\documentclass[12pt]{article}
\usepackage{epsfig} % include eps figures
\usepackage{cite}   % collapse adjacent citations

\newcommand{\beq}{\begin{equation}}
\newcommand{\eeq}{\end{equation}}
\newcommand{\beqs}{\begin{eqnarray}}
\newcommand{\eeqs}{\end{eqnarray}}
\newcommand{\bma}{\begin{matrix}}
\newcommand{\ema}{\end{matrix}}
\newcommand{\bit}{\begin{itemize}}
\newcommand{\eit}{\end{itemize}}
\newcommand{\bay}{\begin{array}}
\newcommand{\eay}{\end{array}}

\newcommand{\bc}{\mbox{\boldmath $c$}}

\newcommand{\bg}{\mbox{\boldmath $g$}}

\newcommand{\bx}{\mbox{\boldmath $x$}}
\newcommand{\bC}{\mbox{\boldmath $C$}}

\newcommand{\bE}{\mbox{\boldmath $E$}}
\newcommand{\bG}{\mbox{\boldmath $G$}}
\newcommand{\bM}{\mbox{\boldmath $M$}}

\newcommand{\bR}{\mbox{\boldmath $R$}}
\newcommand{\bS}{\mbox{\boldmath $S$}}

\newcommand{\IR}{\mbox{$I\!\!R$}}

\newcommand{\hc}{\mbox{\boldmath $\hat{c}$}}

\newcommand{\hg}{\mbox{\boldmath $\hat{g}$}}

\newcommand{\cC}{\mbox{\boldmath ${\cal C}$}}

\newcommand{\andthis}{~~~~~~\mbox{and}~~~~~~}

\def\gsim{\ \rlap{\raise 3pt \hbox{$>$}}{\lower 3pt \hbox{$\sim$}}\ }
\def\lsim{\ \rlap{\raise 3pt \hbox{$<$}}{\lower 3pt \hbox{$\sim$}}\ }

\begin{document}
%%%%%%%%%%%%%%%%

\title{\LARGE The Iterative Signature Algorithm for the analysis of large scale gene expression data}

\author{\\Sven Bergmann, Jan Ihmels and Naama Barkai\footnote{
         Correspondence should be addressed to:
         {\tt Naama.Barkai@weizmann.ac.il}}
         \vspace{5mm} \\
        {\it \small Department of Molecular Genetics,
        Weizmann Institute of Science,
        Rehovot 76100, Israel}}

\date{\vspace{1.0cm} \today}
\maketitle
\vspace{1.5cm}

\begin{abstract}%
We present a new approach for the analysis of genome-wide expression data. Our method is designed to overcome the limitations of traditional techniques, when applied to large-scale data. Rather than alloting each gene to a single cluster, we assign both genes and conditions to context-dependent and potentially overlapping {\it transcription modules}. We provide a rigorous definition of a transcription module as the object to be retrieved from the expression data. An efficient algorithm, that searches for the modules encoded in the data by iteratively refining sets of genes and conditions until they match this definition, is established. Each iteration involves a linear map, induced by the normalized expression matrix, followed by the application of a threshold function. We argue that our method is in fact a generalization of Singular Value Decomposition, which corresponds to the special case where no threshold is applied. We show analytically that for noisy expression data our approach leads to better classification due to the implementation of the threshold. This result is confirmed by numerical analyses based on {\it in-silico} expression data. We discuss briefly results obtained by applying our algorithm to expression data from the yeast {\it S. cerevisiae}. 
\end{abstract}%

%\newpage
%\tableofcontents

\newpage

%=====================
\section{Introduction}
%=====================

DNA microarray experiments monitor the expression levels of thousands of genes simultaneously~\cite{Schena1995,DeRisi1997,Lander1999,Schulze2001}. Using this technology, large sets of genome-wide expression data have been accumulated~\cite{SMD}. For example, the expression levels of the entire yeast genome (comprising $\sim 6200$ genes) have been measured for more than 1000 different experimental conditions~\cite{yeast}. A large number of DNA chip experiments have also been carried out for higher eukaryotes, such as the nematode {\it C. elegans} and the fruit fly {\it Drosophila}, as well as for a variety of both normal and malignant human tissues.

While large scale expression data have the potential to reveal new insights into the transcriptional network that controls gene expression, they also give rise to a major computational challenge: How can one make sense of the massive expression data containing millions of numbers? The classification of the genes and the experimental conditions is an essential first step in reducing the complexity of such data. However, while standard tools, like clustering 
algorithms~\cite{Eisen1998,Spellman1998,Alon1999,Tavazoie1999,Perou1999,Bittner2000,Scherf2000,Staunton2001} (see~\cite{Brazma2000,Altman2001} for reviews) and Singular Value Decomposition (SVD)~\cite{Holter2000,Alter2000}, provide interesting results when applied to relatively small data sets, typically containing tens of experimental conditions and at most several hundred genes, these methods are of limited use for the analysis of large data sets. In particular, a well-recognized drawback of commonly used clustering algorithms is the fact that they assign each gene to a single cluster, while in fact genes that participate in several functions should be included in multiple clusters~\cite{Tomayo1999,Bittner1999,Cheng2000,Getz2000}. Moreover, both in standard clustering methods and SVD,
genes are analyzed based on their expression under {\it all\/} experimental conditions. This is problematic, since cellular processes are usually affected only by a small subset of these conditions, such that most conditions do not contribute relevant information but rather increase the level of background noise. 

In a recent paper~\cite{Ihmels2002} we introduced a new method for the analysis of large-scale gene expression data that was designed to overcome the above-mentioned problems (see Refs.~\cite{Cheng2000,Getz2000} for other recent approaches). A central idea of this work was to integrate prior biological information, like the function or sequence
of known genes, into the analysis of the gene expression data. In the present article we present a complementary method for the analysis of large-scale data that does not require any prior knowledge beyond the expression data. We start by providing a rigorous definition of the type of information we aim to extract from the expression data by introducing the notion of a {\it transcription module} (TM). A TM contains both a set of genes and a set of experimental conditions. The conditions of the TM induce a co-regulated expression of the genes belonging to this TM. That is, the expression profiles of the genes in the TM are the most similar to each other when compared over the conditions of the TM. Conversely, the patterns of gene expression obtained under the conditions of the TM are the most similar to each other when compared only over the genes of the TM. The degree of similarity is determined by a pair of threshold parameters. The gene threshold constrains the gene set, while the condition threshold constrains the condition set. Importantly, distinct transcription modules may share common genes and conditions. 

The precise definition of a TM as the object to be retrieved from the expression data allows us to establish an efficient algorithm that searches for the modules encoded in the data. Starting from a set of randomly selected genes (or conditions) one iteratively refines the genes and conditions until they match the definition of a TM. Using a sufficiently large number of initial sets it is possible to determine all the modules corresponding to a particular pair of thresholds. Scanning through a range of thresholds decomposes the data into modules at different resolutions. 

This paper is organized as follows: In section~\ref{formalism} we provide a mathematical definition of a transcription module.  In section~\ref{ISA} we introduce our algorithm that searches for such modules and compare our method with SVD. In section~\ref{normalization} we discuss the normalization of the expression data. In section~\ref{analysis} we present analytical insight into the role of the threshold in our algorithm. We show that for noisy expression data the application of a threshold improves significantly the identification of transcription modules. We provide an estimate for the maximal amount of noise for which a successful identification is still possible. In section~\ref{beyond-single-module} we compare our method with other standard tools using {\it in-silico} expression data. In section~\ref{application} we discuss briefly results obtained by applying our algorithm to real expression data from the yeast {\it S. cerevisiae}. We conclude in section~\ref{conclusions}.

%==================
\section{Formalism}
%==================
\label{formalism}

%---------------------------------
\subsection{The Expression Matrix}
%---------------------------------

We consider data from microarray experiments given in terms of a gene expression matrix~$\bE$. The matrix element $E^{cg}$ denotes the log-fold expression-change of gene $g \in G \equiv \{1, ..., N_G \}$ at the experimental condition $c \in C \equiv  \{1, ..., N_C \}$, where $N_G$ and $N_C$ refer to the total number of genes and conditions, respectively. The matrix $\bE$ may be viewed as a collection of $N_C$ row vectors:
\beq \label{E-g}
\bE = \pmatrix{\bg_1^T \cr \bg_2^T \cr \vdots \cr \bg_{N_C}^T} \,.
\eeq
Each vector $\bg_c^T = (g_c^{(1)}, g_c^{(2)}, ..., g_c^{(N_G)})$ describes the {\it gene-profile} for condition $c$, containing the expression levels $g_c^{(g)} = E^{cg}$ of all the genes that were monitored under this condition. 
Alternatively the expression matrix can be viewed as a collection of $N_G$ column vectors:
\beq \label{E-c}
\bE = \pmatrix{\bc_1, \bc_2, \dots, \bc_{N_G} } \,.
\eeq
Here each vector $\bc_g = (c_g^{(1)}, c_g^{(2)}, ..., c_g^{(N_C)})^T$ describes the {\it condition-profile} for gene $g$, containing the expression levels $c_g^{(c)} = E^{cg}$ of this gene under all the conditions of the data set. 

We define two normalized expression matrices (c.f. section~\ref{normalization})
\beq \label{norm-E-g}
\bE_G \equiv \pmatrix{\hg_1^T \cr \hg_2^T \cr \vdots \cr \hg_{N_C}^T}  
\eeq
and
\beq \label{norm-E-c}
\bE_C \equiv (\hc_1, \hc_2, ..., \hc_{N_G}) \,.
\eeq
The rows of $\bE_G$ and the columns of $\bE_C$ are given
in terms of the normalized gene- and condition-vectors
\beq \label{normal}
\hg_c \equiv {\bg_c - \bigl< \bg_c \bigr>_{g \in G} \over
 \bigl| \bg_c - \bigl< \bg_c \bigr>_{g \in G} \bigr|} \,,
\andthis
\hc_g \equiv {\bc_g - \bigl< \bc_g \bigr>_{c \in C} \over
 \bigl| \bc_g - \bigl< \bc_g \bigr>_{c \in C} \bigr|}
\eeq
respectively. These vectors have zero mean ($\bigl<\hg_c\bigr>_{g \in G} = \bigl<\hc_g\bigr>_{c \in C} = 0$) and unit length ($|\hg_c| = |\hc_g| = 1$). This normalization implies that $\sum_g {\hat E}_G^{cg} = 0,~\sum_g ({\hat E}_G^{cg})^2 = 1$ for each condition $c$ and $\sum_c {\hat E}_C^{cg} = 0,~\sum_c ({\hat E}_C^{cg})^2 = 1$ for each gene $g$. Centering and re-scaling the rows in $\bE_G$ allows for a meaningful comparison between any two conditions $c$ and $c'$ through their associated gene-profiles $\hg_c$ and $\hg_{c'}$. Similarly, centering and re-scaling the columns in $\bE_C$ allows for the comparison of any two genes $g$ and $g'$ through their associated condition-profiles $\hc_g$ and $\hc_{g'}$. Note that the normalized matrices $\bE_G$ and $\bE_C$ in general are not equal.

%---------------------------------
\subsection{Transcription Modules}
%---------------------------------

Our goal is to find sets of co-regulated genes $G_m \subset G$, together with the relevant experimental conditions $C_m \subset C$ that induce their co-regulation. We refer to such a combined set, $M_m = \{G_m,C_m\}$, as a {\it transcription module} (TM). Here the index $m$ ranges between one and the number of transcription modules, $N_M$. Biologically a TM may be associated with a particular cellular function. Ideally each TM would correspond to a transcription factor that regulates the genes in $G_m$ and that is activated under the conditions in $C_m$. Of course, a one-to-one correspondence between transcription modules and transcription factors is an over-simplification, but it can still provide useful insight into the nature of the expression data. First, the total number of transcription factors, $N_{TF}$, is much smaller than the number of genes: $N_{TF} \ll N_G$. Thus we expect also the number of transcription modules, and therefore the effective dimensionality of the expression matrix to be relatively small: $N_M \ll N_G$. Second, the number of genes activated by a single transcription factor, $N_G^{(m)}$, is known to be limited: $N_G^{(m)} \ll N_G$. Third, different transcription factors can regulate the same gene and can be activated under the same experimental conditions. Hence distinct modules may share common genes and conditions. 

Mathematically a TM can be defined as follows:
\beq \label{consistency}
\exists (T_C, \, T_G):
\left\{
\begin{array}{lll}
C_m(G_m) &=& \Bigl\{c \in C: \bigl<E_G^{cg}\bigr>_{g \in G_m} > T_C \Bigr\} \\ \vspace{-2mm} \\ 
G_m(C_m) &=& \Bigl\{g \in G: \bigl<E_C^{cg}\bigr>_{c \in C_m} > T_G \Bigr\} 
\end{array}
\right. \,,
\eeq
where $T_C$ and $T_G$ are two threshold parameters. The above definition states that for each condition $c$ in the TM the average expression level of the genes in the TM, $\bigl<E_G^{cg}\bigr>_{g \in G_m}$, is above a certain threshold $T_C$. Conversely, for each gene $g$ in the TM the average expression level over the conditions of the TM, $\bigl<E_C^{cg}\bigr>_{c \in C_m}$, is also above some threshold $T_G$. This reciprocal dependence between the genes and the conditions associated with a TM implies that, considering only the genes of the module, the conditions of the module are exactly those for which the co-expression is the most stringent. Similarly, considering only the conditions of the module, the genes of the module are the most tightly co-regulated. Note that our definition of a TM is symmetric with respect to genes and conditions, such that no preference is given to either of them. In particular, we use the expression matrix $\bE_G$ (normalized with respect to genes) in order to specify the conditions of the module ($C_m$), given the genes of the module ($G_m$). Similarly we use $\bE_C$ (normalized with respect to conditions) to specify the genes in $G_m$, given the conditions in $C_m$. 

We would like to reformulate and somewhat generalize the definition of a TM in  eq.~(\ref{consistency}) by introducing vector notation. To this end we represent the genes and the conditions of a TM by a pair of a gene-vector $\bg_m = (g_m^{(1)}, g_m^{(2)}, ...,g_m^{(N_G)})^T$ and a condition-vector $\bc_m = (c_m^{(1)}, c_m^{(2)}, ...,c_m^{(N_C)})^T$. A non-zero component $g_m^{(g)}$ ($c_m^{(c)}$) implies that the gene $g$ (condition $c$) is associated with the module $m$. Consider the linear transformations 
\beq \label{c-proj}
\bc_m^{proj} \equiv \bE_G \, \bg_m = \pmatrix{\hg_1^T \bg_m \cr \hg_2^T \bg_m \cr \vdots \cr \hg_{N_C}^T \bg_m}
\andthis  \label{g-proj}
\bg_m^{proj} \equiv \bE_C^T \bc_m = \pmatrix{\hc_1^T \bc_m \cr \hc_2^T \bc_m \cr \vdots \cr \hc_{N_G}^T \bc_m} \,.
\eeq
The resulting vectors contain the projections of the vectors $\bg_m$ and $\bc_m$, that specify the TM, onto the set of the (normalized) gene-profiles $\{\hg_c\}$ and condition-profiles $\{\hc_g\}$, defined in eq.~(\ref{normal}), that describe the expression data. For a binary vector $\bg_m$ the components of $\bc_m^{proj}$ are just the expression levels summed over the genes of the TM for each condition in the data set. Likewise for a binary vector $\bc_m$ the components of $\bg_m^{proj}$ are the expression levels summed over the conditions of the module for each gene.

The consistency requirement in eq.~(\ref{consistency}) can then be written as
\beq \label {consistency-vec}
\exists (t_C, \, t_G):
\left\{
\begin{array}{lll}
\bc_m &=& f_{t_C}(\bc_m^{proj}) \\
\bg_m &=& f_{t_G}(\bg_m^{proj})
\end{array}
\right. \,,
\eeq
where $t_C$ and $t_G$ are the condition- and the gene-threshold, related to $T_C$ and $T_G$, respectively. The threshold function
\beq \label{threshold-function}
f_t(\bx) \equiv \pmatrix{w(x_1)     \cdot \Theta(\tilde x_1 - t) \cr \vdots \cr 
                         w(x_{N_x}) \cdot \Theta(\tilde x_{N_x} - t)}
\eeq
acts separately on each of the $N_x$ components $x_i$ of the vector $\bx$ and yields the products of a weight-function $w(x)$ and a step-function $\Theta(x)$ as output. The arguments of the step-function, $\tilde x_i = (x_i-\mu(\bx))/\sigma(\bx)$, have been centered and re-scaled. We use the mean as center, $\mu(\bx)=\bigl<\bx\bigr>$, and the expected or measured standard deviation, $\sigma(\bx)=\sqrt{\sum_i^{N_x} (x_i-\bigl<\bx\bigr>)^2/N_x}$, as scale-factor. The step-function sets to zero all elements of the vector $\bx$ that do not exceed $\mu(\bx)$ by at least $t \cdot \sigma(\bx)$. (Down-regulation can be captured by replacing $\tilde x_i \to |\tilde x_i|$ in eq.~(\ref{threshold-function}).) Using $w(x)=1$ as weight-function all the significant elements are set to unity. This binary formulation corresponds to the consistency requirement in eq.~(\ref{consistency}). (To capture down-regulation one uses $\mbox{sign}(x)$ as weight-function.) It is straightforward to extend our formalism using different weight-functions. In this case the entries of the gene- and condition-vector become continuous, and their value determines the significance of a particular gene or condition, respectively. As we shall see, a particularly relevant choice is $w(x)=x$ in which case $f_t(\bx)$ is semi-linear.

The compact definition of a TM in eq.~(\ref{consistency-vec}) can be understood as follows: Applying the threshold function $f_{t_C}$ to $\bc_m^{proj}$ results in a non-zero component $c_m^{(c)}$ of the module's condition-vector $\bc_m$, if the corresponding gene-profile $\hg_c$ is sufficiently aligned with the gene-vector $\bg_m$ of the module. Biologically this means that a significant fraction of the genes in the module are co-regulated under condition $c$. Similarly, the application of $f_{t_G}$ to $\bg_m^{proj}$ results in a non-zero component $g_m^{(g)}$ in the module's gene-vector $\bg_m$, if the corresponding condition-profile $\hc_g$ is sufficiently aligned with the condition-vector $\bc_m$ of the module. Biologically this implies that a significant fraction of the conditions in the module induce a co-regulated expression of gene $g$.

It is important to note that the content of a particular module $M_m = \{G_m, C_m\}$ depends on the pair of thresholds $(t_G, t_C)$. In many cases for slightly larger thresholds there exists a related module $M_m^{up}$, such that $M_m^{up} \subset M_m$. Similarly, for somewhat smaller thresholds there usually exists a module $M_m^{down}$, such that $M_m \subset M_m^{down}$. Thus there are nested sets of modules, $M_m^{top} \subset ... \subset M_m^{bottom}$ that persist over a finite range of the thresholds. This hierarchical structure resembles the tree structures obtained from clustering. However, in our case distinct branches may share common genes or conditions.

%------------------------------------------
\section{The Iterative Signature Algorithm}
\label{ISA}
%------------------------------------------

The rigorous definition of a transcription module in principle allows us to determine the modules encoded in the expression matrix by testing all possible sets $\{G_m,C_m\}$ for their compliance with eq.~(\ref{consistency-vec}). However, since the number of such sets scales exponentially with the number of genes and conditions, such an approach is completely infeasible computationally. We therefore suggest a different approach. Our principle idea is to search for solutions of the consistency equation in~(\ref{consistency-vec}) through the map defined by
\beqs
\bc^{(n+1)} &=& f_{t_C}(\bE_G \, \bg^{(n)}) \,,
\label{cond-signature} \\
\bg^{(n+1)} &=& f_{t_G}(\bE_C^T \, \bc^{(n+1)}) \,.
\label{gene-signature}
\eeqs
The first equation assigns a condition-vector $\bc^{(n+1)}$ to a given gene-vector $\bg^{(n)}$. We refer to the component $c_c^{(n+1)}$ of this vector as a {\it condition score}. This score is non-zero only if the corresponding gene-profile $\hg_c$, defined in eq.~(\ref{normal}), is sufficiently aligned with the gene-vector $\bg_m^{(n)}$. In the subsequent step in eq.~(\ref{gene-signature}) the component (or {\it gene score}) $g_g^{(n+1)}$ of the gene-vector $\bg_m^{(n+1)}$ is assigned a non-zero value only if the corresponding condition-profile $\hc_g$  is sufficiently aligned with the condition-vector $\bc_m^{(n+1)}$.  

In a recent work~\cite{Ihmels2002} we have applied the map in eqs.~(\ref{cond-signature}) and~(\ref{gene-signature}) to a variety of biologically motivated input-sets $\{\bg^{(0)}_i\}$ assembled according to prior knowledge of the regulatory sequence or function of the genes. Sets of co-regulated genes and co-regulating conditions were constructed from recurrent realizations of the output-sets defined by $\bg^{(1)}$ and $\bc^{(1)}$. In this work we pursue a different strategy, namely we apply the maps in eqs.~(\ref{cond-signature}) and~(\ref{gene-signature}) iteratively by re-using the gene-vector $\bg^{(1)}$ as input for eqs.~(\ref{cond-signature}) and~(\ref{gene-signature}) in order to obtain new output-sets defined by $\bc^{(2)}$ and $\bg^{(2)}$. Repeating this procedure we obtain $\{\bg^{(3)},\bc^{(3)}\}$ from $\bg^{(2)}$ and so on. In general, the series $\{\bg^{(0)}, \bg^{(1)}, \bg^{(2)}, \bg^{(3)}, ...\}$ rapidly converges to a ``fixed point'' gene-vector $\bg^{(*)}$. In general the series $\{\bg^{(0)}, \bg^{(1)}, \bg^{(2)}, \bg^{(3)}, ...\}$ rapidly converges and we can define a ``fixed point'' gene-vector $\bg^{(n^*)}$ which satisfies
\beq \label{convergence}
{|\bg^{(*)} - \bg^{(n)}| \over |\bg^{(*)} + \bg^{(n)}|} < \varepsilon 
\eeq
for all $n$ above a certain number of iterations. The parameter $\varepsilon$ determines the accuracy of the fixed point. $\bg^{(*)}$ depends both on the ``seed'' $\bg^{(0)}$ and the thresholds $t_G$ and $t_C$, which are fixed parameters. Together with the associated condition-vector $\bc^{(*)}$ it defines a TM, since $(\bg^{(*)}, \bc^{(*)})$ by definition solve eq.~(\ref{consistency-vec}). We call this procedure the {\it Iterative Signature Algorithm} (ISA).

Although the set of possible input seeds is huge, usually there exist only a rather limited number of fixed points for a given set of thresholds $(t_G, t_C)$. Therefore, in general the ISA is applied as follows: (1)~generate a (sufficiently large) sample of input seeds $\{\bg_m^{(0)}\}$, (2)~find the fixed points $(\bg_m^{(*)},\bc_m^{(*)})$ corresponding to each seed through iterations and (3)~collect the distinct fixed points in order to decompose the expression data into modules. The structure of this decomposition depends on the choice of thresholds $(t_G, t_C)$. Scanning over different values for $(t_G, t_C)$ reveals the modular structure at different resolutions: Lower thresholds yield larger units whose co-regulation is relatively loose, while higher thresholds lead to smaller, tightly co-regulated modules. Each fixed point $(\bg_m^{(*)}, \bc_m^{(*)})$ has its ``basin of attraction'', i.e. the set of seeds that converge to it under the iterative scheme in eqs.~(\ref{cond-signature}) and~(\ref{gene-signature}). The size of this set is a measure of the ``convergence radius'', while the average number of iterations, that is needed until eq.~(\ref{convergence}) is satisfied, characterizes the ``depth'' of this basin.

The computation time of any algorithm, designed for the analysis of large scale expression data, is
of crucial importance. For algorithms that require the full correlation matrices (like clustering or SVD), 
already the computation of these two matrices can be very intensive, since its computation time scales like $t_{comp}^{corr} \propto N_G^2 N_C + N_C^2 N_G$. However, the ISA is not based on this kind of information. Rather than squaring the expression matrix, only multiplications of the expression matrix with {\it sparse} matrices (of size $N_G \times N_I$ or $N_C \times N_I$), where $N_I$ is the number of input sets, have to be performed. Due to the sparseness, the computation time of the ISA goes like $t_{comp}^{ISA} \propto N_{iter} N_I (N_C \tilde N_G + N_G \tilde N_C)$, where $\tilde N_G$ and $\tilde N_C$ refer to the average number of genes and condition, respectively, whose scores are above the threshold, and $N_{iter}$ is the number of iterations until convergence. Thus the computation time of the ISA scales linearly with $N_G$ and $N_C$. In general only very few iterations $N_{iter}$ are needed to find the fixed points. A large number of input sets $N_I$ increases the chances to find the fixed points with a small convergence radius. However, for practical purposes it is useful to accumulate progressively sets a fixed points by running the ISA repeatedly with a moderate value for $N_I$, thus increasing gradually the accuracy of the fixed point decomposition. Importantly, $\tilde N_G$ and $\tilde N_C$ are much smaller than $N_G$ and $N_C$ as long as the respective thresholds are high enough. Finally, we note that $t_{comp}^{ISA}$ could be further improved by choosing the input seeds not completely at random, but using the information of previous runs (e.g. those at a different threshold).

%--------------------------------------------------------
\subsection{Comparison with Singular Value Decomposition}
\label{SVD}
%--------------------------------------------------------

For $w(x)=x$, in the absence of thresholds and neglecting the two different normalizations of the expression data, the iterative scheme reads
\beqs 
\label{find-c}
\hc^{(n)} = {\bE \hg^{(n-1)} \over |\bE \hg^{(n-1)}|} \,, \\
\label{find-g}
\hg^{(n)} = {\bE^T \hc^{(n)} \over |\bE^T \hc^{(n)}|} \,.
\eeqs
The fixed points of the above equations correspond to the pairs of vectors $(\hg_m,\hc_m)$, where $\hg_m = \bg_m/|\bg_m|$ and $\hc_m = \bc_m/|\bc_m|$ are the normalized eigenvectors of $\bE^T \bE$ and $\bE \bE^T$, respectively. Both eigenvectors are associated with the common eigenvalue $\mu_m^2 = |\bE \hg_m|^2 = |\bE^T \hc_m|^2$. It is interesting to note that a Singular Value Decomposition (SVD) of the expression matrix yields exactly those eigenvectors and eigenvalues~\cite{Duda2001,Golub1996} (see appendix~\ref{SVD-appendix} for brief review of SVD). This decomposition is usually performed in a sequential manner. In this case one determines first the pair $(\hg_1,\hc_1)$ associated with the largest eigenvalue $\mu_1^2$. In fact this pair emerges as a fixed point of the above equations for any seed $\bg^{(0)}$ that is not perpendicular to $\hg_1$. It can be shown that the matrix
\beq 
\bE_1 = \mu_1 \hc_1 \, \hg_1^T \,.
\eeq
provides the best rank-1 approximation to $\bE = \bE_1 + \bR_1$, where $\bR_1$ denotes the residual term.
A subsequent diagonalization of $\bR_1$ yields the (orthogonal) pair $(\hg_2,\hc_2)$ associated with the second largest eigenvalue $\mu_2$. Continuing this procedure eventually decomposes the expression matrix into a sum
\beq
\bE = \sum_m^{N_M} \bE_m + \bR_{N_M} 
\eeq
of the rank-1 matrices $\bE_m = \mu_m \hc_m \, \hg_m^T$ with $\mu_m = |\bc_m||\bg_m|$. These matrices can be viewed as a special kind of transcription modules.

One of the advantages of SVD is that the significance of each modular component $\bE_m$ can be determined simply according to the magnitude of the associated eigenvalue. The components associated with small eigenvalues are likely to reveal no real information and to contain only noise. Thus the spectrum of eigenvalues can give some indication of the dimensionality of the data: The existence of $N_M$ eigenvalues that are significantly larger than the remaining eigenvalues suggests that there are $N_M$ dominant components. Similar to SVD the lengths of the fixed point vectors of the ISA provide a measure of the relative importance of the associated TM. Specifically, $|\bg_m^{(*)}|^2 = \sum_{g \in G_m} (g^{(*)}_g)^2$ reflects the size of the gene set and (for $w(x)=x$) the strength of its co-regulation, while $|\bc_m^{(*)}|^2 = \sum_{c \in C_m} (c^{(*)}_c)^2$ reflects the size of the condition set and the strength of the co-regulation induced by this set. 

While the similarity between the ISA and SVD is instructive, there are several important differences: 
\bit
\item Applying the threshold functions in eqs.~(\ref{cond-signature}) and~(\ref{gene-signature}) yields a different spectrum of fixed points: Sets of genes that are fixed points of the iterative scheme for a particular choice of the threshold, in general do not correspond to the eigenvectors of the expression matrix. 
\item The thresholds affect the stability of the fixed points: While the iterations in eqs.~(\ref{find-c}) and~(\ref{find-g}) have only a single {\it stable} fixed point $(\hg_1,\hc_1)$, the ISA in eqs.~(\ref{cond-signature}) and~(\ref{gene-signature}) usually possesses several stable fixed points. This is essentially because the thresholds induce an ``effective orthogonality'' by setting the small scalar products in eq.~(\ref{c-proj}) to zero. Consequently input sets that are almost (but not exactly) orthogonal to the strongest fixed point, do not flow towards this point under the iterations, but converge to a different fixed-point.
\item SVD is very sensitive to the (unavoidable) noise in the expression data. This noise induces mixing between modules that would be orthogonal to each other in the absence of noise. In the ISA the threshold function provides an efficient way to deal with such noise. Excluding the bulk of the genes and conditions from the expression data at each step of the iterative procedure allows to pick up co-regulated units that would otherwise be masked by the noise. 
\item For SVD distinct eigenvectors $\hg_m$ and $\hg_{m'}$ as well as $\hc_m$ and $\hc_{m'}$ are orthogonal to each other, since they diagonalize a symmetric matrix. The constraint of orthogonality is not present in the ISA. 
\item SVD only reveals one single decomposition of the expression matrix into modules. As for the ISA, changing the values of the thresholds allows to analyze the modular structure recorded in the expression matrix at different resolutions. 
\item For SVD the expression data has to be normalized either according to genes or conditions. The choice of data normalization in general follows from the interpretation of the data. Demanding maximal variance among the principal components, one is led to center the data either as in $\bE_G$ or $\bE_C$ (see appendix~\ref{SVD-appendix} on SVD for details). Thus the symmetry between the genes and the conditions is explicitly broken when committing to either $\bE_C$ or $\bE_G$. In contrast, the ISA avoids this bias by alternating between the two possible normalizations at each step of the iterative procedure in eqs.~(\ref{cond-signature}) and~(\ref{gene-signature}). 
\eit
We will discuss now some of these points in more detail.

%--------------------------------------
\section{The proper data normalization}
\label{normalization}
%--------------------------------------

Given the ``raw'' expression data contained it is difficult to compare two experiments ($\bg_c$ and $\bg_{c'}$) or two genes ($\bc_g$ and $\bc_{g'}$). This is because different experiments may affect the expression levels at a different scale. For example one condition may change the expression of many genes by a very large factor ($\gg 1$) while another condition affects mainly the same genes, but shifts their expression level by a much smaller amount. Although the two conditions are related, this relation is not explicit in the expression data. Moreover, recording the expression levels with different microarray techniques as well as variations in the sample preparation can change the scale of the results. Similarly the dynamic range of two distinct genes could differ greatly even though the shape of their condition profiles might be similar. To overcome this difficulty we have introduced the normalized matrices $\bE_G$ and $\bE_C$ (c.f. eqs.~(\ref{norm-E-g}) and~(\ref{norm-E-c})). 

\begin{figure}[\figpos]
\begin{center}
  \mbox{\epsfig{figure=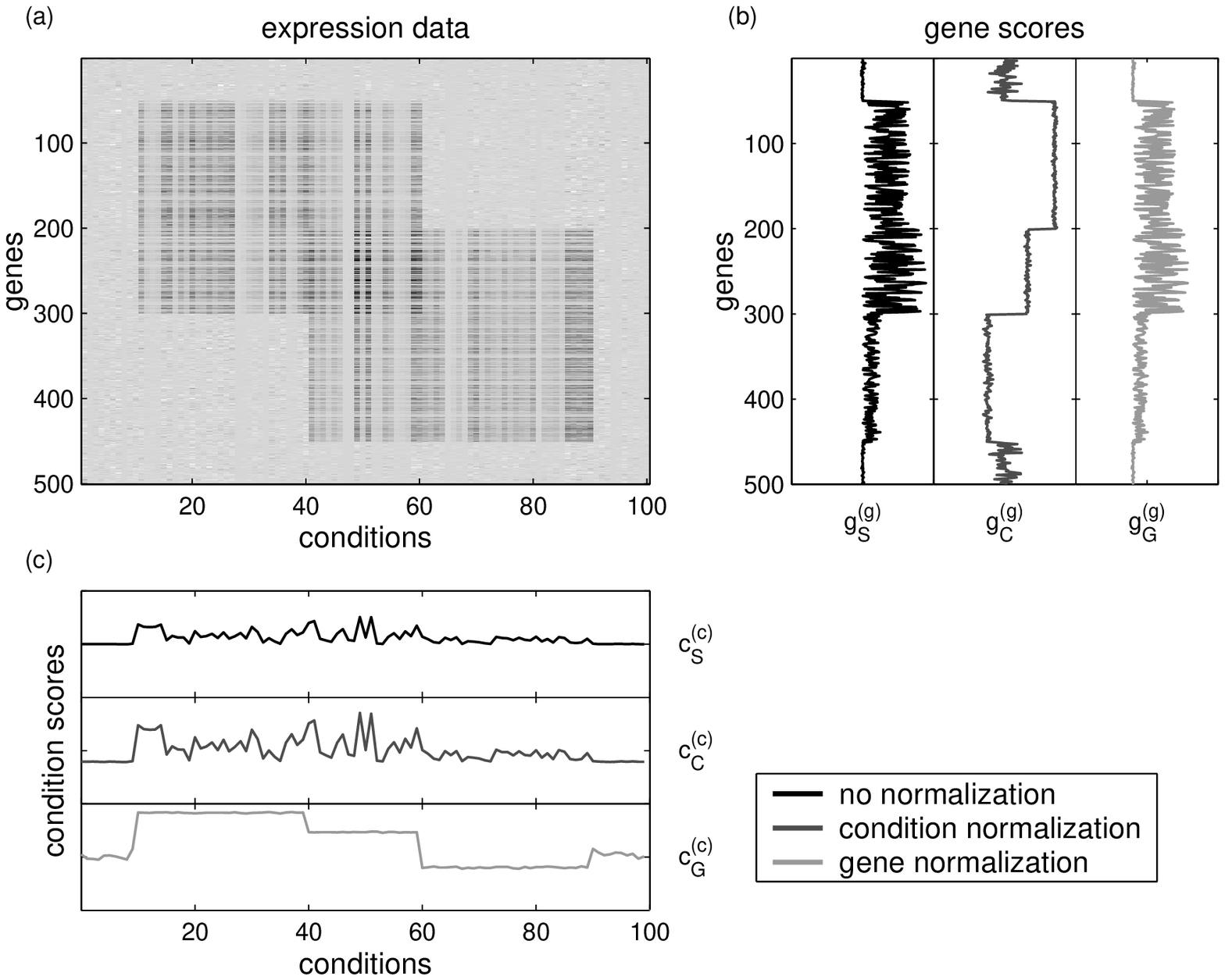,angle=0,width=13cm,height=10cm}}
\end{center}
\vspace{1cm} \caption{{\it How to properly normalize the expression matrix.} 
(a)~An {\it in-silico} expression matrix, corresponding to two overlapping modules of equal size and strength, was generated according to the model described in the text. The elements of the original expression matrix $E^{cg}$, were scaled to $E^{cg}_S \equiv E^{cg} s_g s_c$, where $s_g \in [0,1]$ and $s_c \in [0,1]$ are random scale factors selected from a uniform distribution for each gene~$g$ and condition~$c$. From $\bE_S$ we calculated the normalized expression matrices $\bE_G$ and $\bE_C$ according to eqs.~(\ref{norm-E-g}) and~(\ref{norm-E-c}).
(b)~From the vector $\bc_1$, whose non-zero components $c_1^{(c)}$ specify the conditions of the upper-left module in (a) we calculated the vectors $\bg_S = \bE_S^T \bc_1$, $\bg_C = \bE_C^T \bc_1$ and $\bg_G = \bE_G^T \bc_1$. We
plot their components (horizontal axes) $g_S^{(g)}$ (black), $g_C^{(g)}$ (dark gray) and $g_G^{(g)}$ (light gray) as a function of the gene index (vertical axis). Only for $\bg_C$, obtained according to normalization used in the ISA, {\it all} the components associated with the genes of the module are significantly larger than the others.
(c)~From the vector $\bg_1$, whose non-zero components $g_1^{(g)}$ specify the genes of the upper-left module in (a) we calculated the vectors $\bc_S = \bE_S \bg_1$, $\bc_C = \bE_C \bg_1$ and $\bg_G = \bE_G \bc_1$. We plot their components (horizontal axes) $c_S^{(c)}$ (black), $c_C^{(c)}$ (dark gray) and $c_G^{(c)}$ (light gray) as a function of the condition index (vertical axis). Only for $\bc_G$, obtained according to normalization used in the ISA, {\it all} the components associated with the conditions of the module are significantly larger than the others.}
\label{normalizations}
\end{figure}

In order to study the impact of the normalization on our algorithm we generated an {\it in-silico} expression matrix $\bE$ corresponding to two overlapping modules of equal size and strength (see section~\ref{beyond-single-module} for more details on the model used to generate these data). We selected random scale factors $s_g, s_c \in [0,1]$ for each gene $g$ and condition $c$ from a uniform distribution and transformed the elements of the expression matrix according to $E^{cg} \to E^{cg}_S \equiv E^{cg} s_g s_c$. Unlike the original expression matrix $\bE$, the re-scaled expression matrix $\bE_S$ (shown in Fig.~\ref{normalizations}a) corresponds to the realistic scenario where the entities of the expression data have been recorded at different scales. From $\bE_S$ we calculated the normalized matrices $\bE_C$ and $\bE_G$.  

The question we ask is which normalization has to be employed in order to reveal the ``correct'' genes from the conditions associated with the underlying module, and which normalization leads to the ``correct'' conditions, given the genes of the module. To answer this question we defined the vectors $\bg_1$ and $\bc_1$ by assigning non-zero components only for the genes and conditions of one of the modules, respectively. Using these vectors we computed \linebreak $\bc_S = \bE_S \bg_1$, $\bc_C = \bE_C \bg_1$ and $\bc_G = \bE_G \bg_1$ as well as $\bg_S = \bE_S^T \bc_1$, $\bg_C = \bE_C^T \bc_1$ and $\bg_G = \bE_G^T \bc_1$. The components of the resulting gene- and condition-vectors are plotted in Fig.~\ref{normalizations}b and~c, respectively. 

One can see that only for $\bg_C$ and $\bc_G$ (corresponding to the the ``correct'' normalizations as used in the ISA, c.f. eqs.~(\ref{cond-signature}) and~(\ref{gene-signature})) {\it all} the components associated with the genes and conditions of the module (specified by $(\bg_1, \bc_1)$) are significantly larger than the others. For missing or ``wrong'' normalization there are large fluctuations among the vector components. Hence applying a threshold would only capture part of the relevant genes or conditions in this case. Thus $\bE_C$ is best suited to identify the genes of a module from a set of conditions that is a good approximation of $C_m$, while $\bE_G$ is the proper normalization to obtain the conditions of a module from a set of genes close to $G_m$. Note that using these ``correct'' normalizations, it is even possible to distinguish the genes and conditions associated exclusively with the specified module from those that belong also to the other module, because the latter obtain a somewhat lower score.

%============================
\section{Analysis of the ISA} 
\label{analysis}
%============================

The fundamental issue is how well the ISA can reveal relatively small, noisy, and possibly overlapping modules from the expression matrix. In this section we address this question by considering a simple model where the expression matrix corresponds to a {\it single} transcription module. Our idea is to consider the gene-vector that undergoes iterations as a stochastic entity and to study how its distribution evolves under the iterations. This approach allows us to quantify how the efficiency of our algorithm depends on the size of the module and the noise in the expression data.

%-----------------------------
\subsection{Linear recursions} 
%-----------------------------

In the following we consider a slightly simplified iterative scheme, where no threshold function is applied to the condition vector. In this case one can write an iterative equation that depends only on the gene vector.
If, moreover, no gene threshold is applied the iterations are defined through the linear equation (c.f. eq.~(\ref{eigen-genes}) in the Appendix)
\beq 
\label{iterate-g}
\hg^{(n)} = {\cC \bg^{(n-1)} \over |\cC \bg^{(n-1)}|} \,.
\eeq
Here the matrix $\cC = \bE^T \bE$ emerges from applying first eq.~(\ref{find-c}) and then eq.~(\ref{find-g}). As we mentioned before the fixed points of this linear recursion are the eigenvectors of $\cC$.

Let us consider the simplest scenario corresponding to a single set of co-regulated genes $G_1 \subset G$ whose co-regulation is triggered by the conditions in $C_1 \subset C$. Specifically, we assume that all the genes in $G_1$ are equally important, such that a noise-free measurement would result in identical condition profiles for these genes. In this ideal case the matrix elements ${\cal C}^{gg'}$ would equal some constant if both $g$ and $g'$ belong to $G_1$ and be zero otherwise. In order to model the effect of noisy data we consider the elements of $\cC$ as random variables with mean value 
\beq \label{mean-C}
\bigl<{\cal C}^{gg'}\bigr> = 
\left\{\matrix{\mu_{\cal C} & g,g' \in G_1 \cr 0 & \mbox{otherwise}}\right. \,,
\eeq
and variance $V_{\cal C}$ for all $g,g' \in G$. In the absence of noise (i.e.~$V_{\cal C}=0$) the matrix $\cC$ possesses only a single (non-trivial) eigenvector $\bg^{(0)}$, whose non-zero components specify the genes of the TM. However, for $V_{\cal C}>0$ this is not true anymore. 

Assume we knew the eigenvector of $\cC$ for $V_{\cal C}=0$ and use it as a (binary) seed $\bg^{(0)}$ for eq.~(\ref{iterate-g}) with a noisy realization of $\cC$ (i.e. $V_{\cal C}>0)$. The question is whether the fixed-point resulting from $\bg^{(0)}$ still characterizes the genes of the module. In general the vector $\hg^{(1)}$ obtained by the first iteration does not coincide with $\hg^{(0)}$. Due to the probabilistic description of $\cC$ we can only determine the mean and the variance of the components of $\bg^{(1)} = \cC \bg^{(0)}$. The mean of $g_g^{(1)} = \sum_{g'} {\cal C}^{gg'} g_{g'}^{(0)}$ is equal to the number of genes in the module, $N_G^{(m)}$, times $\mu_{\cal C}$ if $g \in G_1$, and zero otherwise. Similarly the variance of $g_g^{(1)}$ is $N_G^{(m)} V_{\cal C}$. Here we only used the additivity of the mean and the variance. However, already for $g_g^{(2)}$ in the next iteration we need to deal with products of random variables. To this end we note that for two independent random variables $a$ and $b$ we have (see appendix~\ref{variance-product} for proof)
\beq \label{multi-mean-var}
\bigl<a b\bigr> = \bigl<a\bigr> \bigl<b\bigr> \andthis 
V(a b) = V(a) \, V(b) + V(a) \, \bigl<b\bigr>^2 +  V(b) \, \bigl<a\bigr>^2 \,.
\eeq
Using these results we find that the mean values of the components of the vector $\bg^{(n)} = \cC \bg^{(n-1)}$ are given by 
\beq \label{mean-g}
\bigl<g^{(n)}_g\bigr> = 
\left\{\matrix{\mu_G^{(n)} \equiv N_G^{(m)} \mu_{\cal C} \, \mu_G^{(n-1)} & g \in G_1 \cr 0 & g \not \in G_1} \right. \,,
\eeq
where $\mu_G^{(n-1)}$ denotes the mean of the components $g^{(n-1)}_g$ associated with the module ($g \in G_1$). Only for the genes in $G_1$ there are $N_G^{(m)}$ matrix elements in $\cC$ that contribute constructively to $\bigl<g^{(n)}_g\bigr>$. Similarly, the variances of $g^{(n)}_g$ are
\beq \label{var-g}
V(g^{(n)}_g) = 
\left\{
\begin{array}{ll} 
V_G^{(n)} \equiv \Delta N_G V_{\cal C} \tilde V_G^{(n-1)} + 
N_G^{(m)} \left(V_{\cal C} V_G^{(n-1)} + V_{\cal C} (\mu_G^{(n-1)})^2 + V_G^{(n-1)} \mu_{\cal C}^2 \right) & 
g \in G_1 \\ 
\tilde V_G^{(n)} \equiv \Delta N_G V_{\cal C} \tilde V_G^{(n-1)} + 
N_G^{(m)} V_{\cal C} \left(V_G^{(n-1)} + (\mu_G^{(n-1)})^2 \right) & 
g \not \in G_1 
\end{array} 
\right. \,,
\eeq
where $\Delta N_G \equiv N_G - N_G^{(m)}$ denotes the number of genes that do not belong to the module. Note that $V_G^{(n)}$ has an additional term with respect to $\tilde V_G^{(n)}$, due to the contribution of the non-zero mean values in $\cC$.

In order to assess whether the iterations improve the separability between distributions of the genes within ($g \in G_1$) and outside ($g \not \in G_1$) the module, we introduce the re-scaled variances
\beq
v_G^{(n)} \equiv {V_G^{(n)} \over (\mu_G^{(n)})^2} \andthis
\tilde v_G^{(n)} \equiv {\tilde V_G^{(n)} \over (\mu_G^{(n)})^2} \,. 
\eeq
Note that $v_G^{(n)}$ and $\tilde v_G^{(n)}$ are dimensionless and invariant under the normalization of the gene-vectors. $v_G^{(n)} \ll 1$ implies that the distribution of the genes associated with the module is well separated from the distribution of the genes that do not belong to the module. Using eqs.~(\ref{mean-g}) and~(\ref{var-g}) we obtain the following recursive equations 
\beqs \label{v-G-n-tilde}
\tilde v_G^{(n)} &=& {\Delta N_G v_{\cal C} \over (N_G^{(m)})^2} \tilde v_G^{(n-1)} + 
{v_{\cal C} \over N_G^{(m)}} \left(v_G^{(n-1)} + 1 \right) \,, \\
\label{v-G-n}
v_G^{(n)} &=& \tilde v_G^{(n)} + {v_G^{(n-1)} \over N_G^{(m)}} \,, 
\eeqs
where $v_{\cal C} \equiv {V_{\cal C} / \mu_{\cal C}^2}$ is the (fixed) noise-to-signal ratio of the expression matrix.

If $N_G^{(m)} \gg 1$ the second term in eq.~(\ref{v-G-n}) is negligible and we can ignore the small difference between $v_G^{(n)}$ and $\tilde v_G^{(n)}$. Then, setting $\tilde v_G^{(n)} = v_G^{(n)}$ in eq.~(\ref{v-G-n-tilde}) leads to the approximate recursive equation
\beq \label{approximate-recursions}
v_G^{(n)} = {N_G v_{\cal C} \over (N_G^{(m)})^2}  v_G^{(n-1)} + 
{v_{\cal C} \over N_G^{(m)}} \,. \\
\eeq
This equation converges to
\beq \label{v-fixed-point}
v_G^{(*)} \equiv \left({N_G^{(m)} \over v_{\cal C}} - {N_G \over N_G^{(m)}} \right)^{-1} \,,
\eeq
provided that
\beq \label{v-C-critical}
v_{\cal C} < v_{\cal C}^{crit} \equiv {(N_G^{(m)})^2 \over N_G} \,.
\eeq
For further reference we state this result also for the {\it signal-to-noise ratio}
\beq
\rho_G^{(n)} \equiv {\mu_G^{(n)} \over \sqrt{V_G^{(n)}}} = (v_G^{(n)})^{-1/2} \,.
\eeq
The corresponding fixed-point value equals to
\beq \label{fixed-point-rho}
\rho^{(*)}_G =  
\left[N_G^{(m)} \left(\rho_{\cal C}^2 - (\rho_{\cal C}^{crit})^2 \right) \right]^{1/2} \,,
\eeq
if 
\beq \label{rho-C-critical}
\rho_{\cal C} \equiv {\mu_{\cal C} \over \sqrt{\sigma_{\cal C}}} >
 \rho_{\cal C}^{crit} \equiv {\sqrt{N_G} \over N_G^{(m)}} \,,
\eeq
and is zero otherwise.

The interpretation of the critical value $v_{\cal C}^{crit}$ for the noise in the expression data is straightforward: Only sets of genes that are sufficiently large and whose co-regulation is recorded in the expression matrix with relatively low noise (i.e. $v_{\cal C} < v_{\cal C}^{crit}$) can be captured by the iterative procedure without threshold in eq.~(\ref{iterate-g}). Actually eq.~(\ref{rho-C-critical}) is only a necessary condition for the identification of a module, since for a reliable separation of the distributions of the gene-scores associated with the module, we need $\rho_G^{(*)} \gg 0$. As we mentioned before, the number of genes associated with cellular functions is expected to be rather limited, $N_G^{(m)} \ll N_G$. Therefore we conclude that eq.~(\ref{rho-C-critical}) presents a serious limitation for the extraction of biologically relevant modules through the analysis of the eigenvectors of~$\cC$ (as in SVD).

%-----------------------------------------------------
\subsection{Noise reduction by the threshold function} 
%-----------------------------------------------------

As discussed in the previous section the noise in the expression data may obstruct the identification of a TM. A fundamental aspect of the threshold functions in the ISA is to reduce the effect of such noise by excluding the bulk of the genes and conditions that do not contribute information but rather increase the level of background noise. 

To illustrate this point, let us repeat the study of noise propagation presented above for the simplified iterative scheme like in eq.~(\ref{iterate-g}), but with the linear map followed by a threshold function:
\beq 
\label{iterate-g-threshold}
\bg^{(n)} = f_t(\cC \hg^{(n-1)}) \,,
\eeq
where $f_t$ is defined in eq.~(\ref{threshold-function}) and we use a linear weight-function $w(x)=x$.
Let us assume that the gene scores are distributed according to normal distributions ${\cal N}(x; \mu, \sigma)$, where
$\mu$ and $\sigma$ refer to the mean and the standard deviation of the random variable $x$. 
As a result of the threshold function only 
\beq
\label{N-G-eff}
\tilde N_G^{(m)} = N_G^{(m)} \, \int_t^{\infty} {\cal N}(\rho; \rho_G^{(n-1)}, 1) \, d\rho 
\eeq
genes from the module contribute constructively to the mean in eq.~(\ref{mean-g}). Similarly, only
$\tilde N_G^{(m)}$ genes from the module and 
\beq
\label{delN-eff}
\Delta \tilde N_G = \Delta N_G \, \int_t^{\infty} {\cal N}(\rho; 0, 1) \, d\rho 
\eeq
genes outside the module contribute to the variance of $g_g^{(n)}$ in eq.~(\ref{var-g}). $\tilde N_G^{(m)}$ is the expected number of genes in the module, whose score has not been set to zero by the threshold function. Similarly, $\Delta \tilde N_G$ is the expected number of genes that do not belong to the module, but have a non-zero score. The crucial point is that, because of the different mean values of the two distributions, the threshold function excludes more genes that do not belong to the module than genes that do belong to the module. For example, if $\rho_G^{(0)}=3$ for the initial (normal) distribution, then a threshold $t=2$ would remove almost 98\% of the genes outside the module ($\Delta \tilde N_G \simeq 0.023 \times \Delta N_G$), but less than 16\% of the genes associated with the module ($\tilde N_G^{(m)} \simeq 0.841 \times N_G^{(m)}$). We note that the precise shape of the distribution function is in fact not crucial, since our derivation relies only on the additivity of the mean values and variances, and eq.~(\ref{multi-mean-var}).

\begin{figure}[\figpos]
\begin{center}
  \mbox{\epsfig{figure=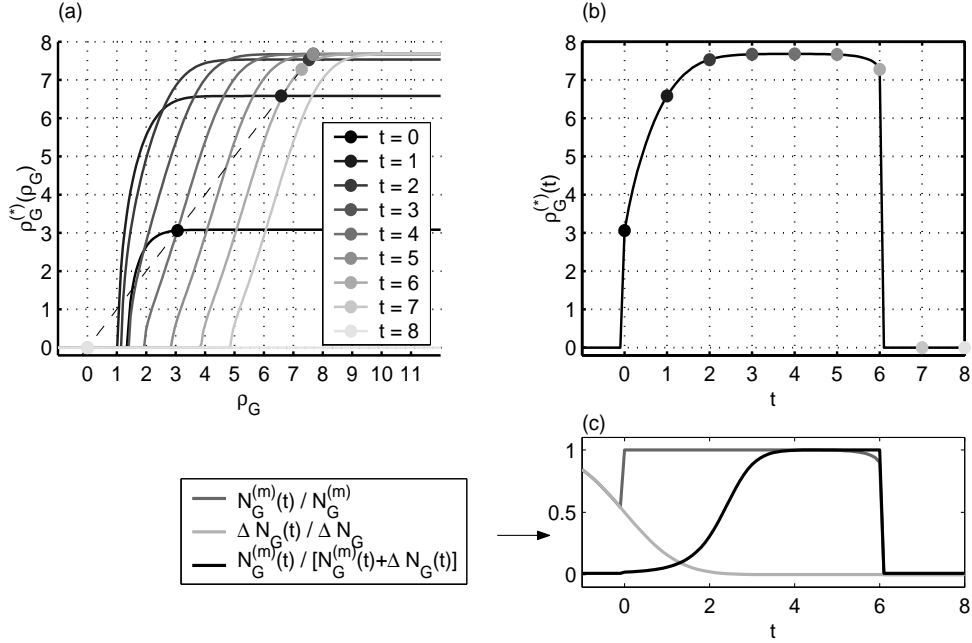,angle=0,width=13cm,height=8.7cm}}
\end{center}
\vspace{0.0cm} 
\caption{{\it Finding the fixed point value of the signal-to-noise ratio.} (a)~The fixed point value of the signal-to-noise ratio $\rho_G^{(*)}(t)$ is found by solving eq.~(\ref{fixed-point-rho-threshold}) (c.f. section~\ref{analysis}). We plot its right-hand-side $RHS(\rho_G,t) \equiv \left[\tilde N_G^{(m)} \rho_{\cal C}^2 - (\tilde N_G^{(m)} + \Delta \tilde N_G)/\tilde N_G^{(m)} \right]^{1/2}$ as a function of $\rho_G$ for several values of the threshold $t$ as indicated in the legend (setting $N_G=6000$, $N_G^{(m)}=60$, $\rho_{\cal C} = 1$). $RHS(\rho_G,t)$ depends on $\rho_G$ and $t$ through the effective numbers $\tilde N_G^{(m)}(t,\rho_G)$ and $\Delta \tilde N_G(t)$ (defined in eqs.~(\ref{N-G-eff}) and~(\ref{delN-eff})) that denote the expected number of genes inside and outside the module that passed the threshold. Each curve increases monotonically from zero to its maximal value $\rho_G^{max}(t)$. For $\rho_G \gg t$, the effective number  $\tilde N_G^{(m)}$ approaches $N_G^{(m)}$. In this limit $\rho_G^{max}(t)$ depends on $t$ only through $\Delta \tilde N_g$, which goes to zero for $t \gg 1$. Thus $\rho_G^{max}(t) \to \sqrt{N_G^{(m)} \rho_{\cal C}^2 - 1}$ asymptotically. According to eq.~(\ref{fixed-point-rho-threshold}) the fixed-point solutions for the signal-to-noise ratio $\rho_G^{(*)}(t)$ are given by $\rho_G = RHS(\rho_G,t)$ and therefore correspond to the intersections (indicated by the big dots) of these curves with the diagonal (shown as a dashed line). (b)~The solutions in (a) are plotted as a function of the threshold $t$. For a relatively small threshold ($t \lsim 2$) $\rho_G^{(*)}(\rho_G,t)$ increases rapidly as a function of $t$, saturates to $\rho_G^{max}$ for $t \gsim 2$ and suddenly falls off to zero at a certain threshold $t_{trans} (\approx 6)$. This behavior can be understood from (a): For a low threshold the intersection of curves for $RHS(\rho_G,t)$ with the diagonal appears at small values of $\rho_G$. For larger $t$ the intersections occur in the saturated regime of $RHS(\rho_G,t)$, such that $\rho_G \simeq \rho_G^{max}(t)$. However, if $t$ is too large the curves do not intersect with the diagonal and there is no solution. (c)~$\tilde N_G^{(m)}(t) / N_G^{(m)}$ (dark gray) as well as  $\Delta \tilde N_G(t)/\Delta N_G$ (light gray) and $\varrho(t) \equiv \tilde N_G^{(m)}(t)/(\tilde N_G^{(m)}(t)+\Delta \tilde N_G(t))$ (black) are shown as a function of $t$. $\varrho(t) \simeq 1$ for $3 \lsim t < 6$, indicating the optimal regime for the threshold.}
\label{single-module-2d}
\end{figure}

It follows that the mean values and variances of the components of the vector $\bg^{(n)}$ are given by the same expression as in eqs.~(\ref{mean-g}) and~(\ref{var-g}), respectively, except that we have to replace $N_G^{(m)}$ by $\tilde N_G^{(m)}$ and $\Delta N_G$ by $\Delta \tilde N_G$. Substituting the effective numbers $\tilde N_G^{(m)}$ and $\Delta \tilde N_G$ into eqs.~(\ref{mean-g}) and~(\ref{var-g}) the argument leading to the expression for the fixed-point signal-to-noise ratio in eq.~(\ref{fixed-point-rho}) is essentially unchanged, and we have 
\beq \label{fixed-point-rho-threshold}
\rho^{(*)}_G =  
\left[\tilde N_G^{(m)} \left(\rho_{\cal C}^2 - (\tilde \rho_{\cal C}^{crit})^2 \right) \right]^{1/2} \,,
\eeq
with
\beq \label{rho-C-critical-threshold}
\tilde \rho_{\cal C}^{crit} \equiv {\sqrt{\tilde N_G^{(m)} + \Delta \tilde N_G} \over \tilde N_G^{(m)}} \,.
\eeq
Note that unlike for eq.~(\ref{fixed-point-rho}), the right-hand side of eq.~(\ref{fixed-point-rho-threshold}) still depends on $\rho_G^{(*)}$ through $\tilde N_G^{(m)}$. Therefore eq.~(\ref{fixed-point-rho-threshold}) is an integral equation for $\rho_G^{(*)}$ which can be solved numerically. A graphical solution of this equation is provided in Fig.~\ref{single-module-2d} for different thresholds and a specific choice of the parameters
$N_G$, $N_G^{(m)}$ and $v_{\cal C}$ (see caption for details). 

\begin{figure}[\figpos]
\begin{center}
  \mbox{\epsfig{figure=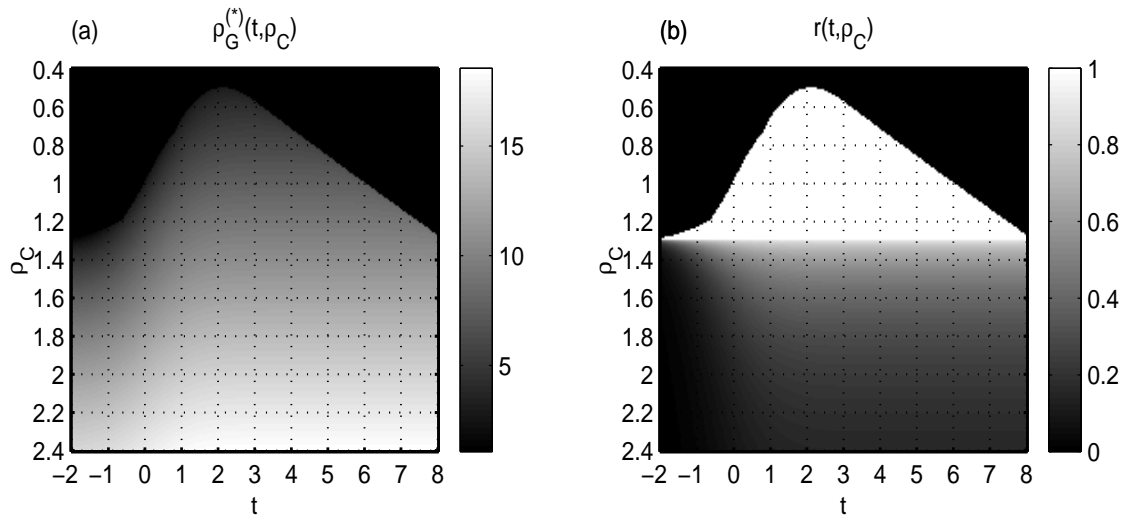,angle=0,width=15cm,height=7cm}}
\end{center}
\vspace{1cm} \caption{{\it Properties of the fixed point value of the signal-to-noise ratio.} (a) The fixed point value of the signal-to-noise ratio, $\rho_G^{(*)}(t, \rho_{\cal C})$, characterizes the separability between the gene score distributions for the genes inside and outside the single module (c.f. section~\ref{analysis} for details). The plot shows  $\rho_G^{(*)}(t, \rho_{\cal C})$ as a function of both the threshold $t$ and the (fixed) signal-to-noise ratio in the expression matrix $\rho_{\cal C}$. For very small thresholds $\rho_G^{(*)}(t, \rho_{\cal C})$ vanishes if $\rho_{\cal C}$ is below some critical value $\rho_{\cal C}^{crit} \approx 1.3$. However, increasing the threshold the iterations converge to a finite fixed-point, $\rho_G^{(*)}(t, \rho_{\cal C}) >0$, even if $\rho_{\cal C} < \rho_{\cal C}^{crit}$ (but $\rho_{\cal C} > \tilde \rho_{\cal C}^{crit} \gsim 0.5$). There is an optimal regime for the threshold $t$, where $\rho_G^{(*)}(t, \rho_{\cal C})$ is (near to) maximal. Within this regime $\rho_G^{(*)}(t, \rho_{\cal C})$ depends only weakly on $t$, so the convergence is robust with respect to the exact choice of the threshold. The size of this regime increases with $\rho_{\cal C}$. (b)~The ratio $r(t, \rho_{\cal C}) \equiv (\rho_G^{(*)}(t, \rho_{\cal C}) - \rho_G^{(*)}(\rho_{\cal C}))/ \rho_G^{(*)}(t, \rho_{\cal C})$ characterizes the improvement in the identification of transcription modules that is achieved by the application of the threshold function. ($\rho_G^{(*)}(\rho_{\cal C})$ denotes the fixed-point value of the signal-to-noise ratio in the absence of a threshold, and $r(t, \rho_{\cal C})$ is set to zero for $\rho_G^{(*)}(t, \rho_{\cal C}) = 0$.) We show $r(t, \rho_{\cal C})$ as function of $t$ and $\rho_{\cal C}$. The regime where $\rho_{\cal C} < \rho_{\cal_C}^{crit}$ is subdivided into a white region ($r(t, \rho_{\cal C}) = 1$), where the iterative scheme only converges to a positive value, $\rho_G^{(*)}(t, \rho_{\cal C}) >0$, due to the threshold and a black area ($r(t, \rho_{\cal C}) = 0$), where the iterative schemes does not converge to a positive value implying that the module cannot be identified in this regime. Note that also for $\rho_{\cal C} > \rho_{\cal C}^{crit}$, where the iterative schemes converges to a positive value even without a threshold, there exists a large region in the parameter space of $t$ and $\rho_{\cal C}$ (the light gray area for $r(t, \rho_{\cal C})$), where $\rho_G^{(*)}(t, \rho_{\cal C})$ is significantly larger than $\rho_G^{(*)}(t)$.} 
\label{single-module-3d}
\end{figure}

As can be seen in Fig.~\ref{single-module-3d}a applying a threshold function improves significantly the identification of the module. We show the fixed point value of the signal-to-noise ratio, $\rho^{(*)}_G$, as a function of both the threshold $t$ and the (fixed) signal-to-noise ratio $\rho_{\cal C}$ of the expression data. In the absence of a threshold function $\rho_G^{(n)}$ converges to zero if $\rho_{\cal C}$ is below some critical value $\rho_{\cal C}^{crit}$. Applying a threshold, $\rho_G^{(n)}$ converges to a finite value, even if $\rho_{\cal C} < \rho_{\cal C}^{crit}$ (but $\rho_{\cal C} > \tilde \rho_{\cal C}^{crit}$), indicating the identification of the module. Moreover, one can see from Fig.~\ref{single-module-3d}a that there is an optimal regime for the threshold $t$, where $\rho_G^{(*)}(t, \rho_{\cal C})$ is (nearly) maximal. Within this regime $\rho_G^{(*)}(t, \rho_{\cal C})$ depends only weakly on $t$, so the convergence is robust with respect to the exact choice of the threshold. The size of this regime increases with $\rho_{\cal C}$. 

In order to quantify the relative increase of the fixed point value of the signal-to-noise ratio $\rho_G^{(*)}(t, \rho_{\cal C})$ due to the application of the threshold function we define the ratio
\beq
r(t, \rho_{\cal C}) \equiv 
{\rho_G^{(*)}(t, \rho_{\cal C}) - \rho_G^{(*)}(\rho_{\cal C}) \over \rho_G^{(*)}(t, \rho_{\cal C})} \,,
\eeq
where $\rho_G^{(*)}(\rho_{\cal C})$ refers to the value to which the signal-to-noise ratio converges when no threshold is applied. For $\rho_G^{(*)}(t, \rho_{\cal C}) = 0$ we set $r(t, \rho_{\cal C})$ to zero. We show $r(t, \rho_{\cal C})$ as a function of $t$ and $\rho_{\cal C}$ in Fig.~\ref{single-module-3d}b. The figure shows 
that there exists a large region in the parameter space of $t$ and $\rho_{\cal C} < \rho_{\cal_C}^{crit}$, where the iterations only converge to a positive value due to the threshold. Moreover, even for $\rho_{\cal C} > \rho_{\cal C}^{crit}$, where the iterative schemes converges to a positive value also without a threshold, there exists a large region, where $\rho_G^{(*)}(t, \rho_{\cal C})$ is significantly larger than $\rho_G^{(*)}(t)$. 
Thus we conclude that the threshold function improves significantly (and in certain cases makes at all possible) the convergence of a noisy input set to a gene-vector that specifies the TM. 

\begin{figure}[\figpos]
\begin{center}
  \mbox{\epsfig{figure=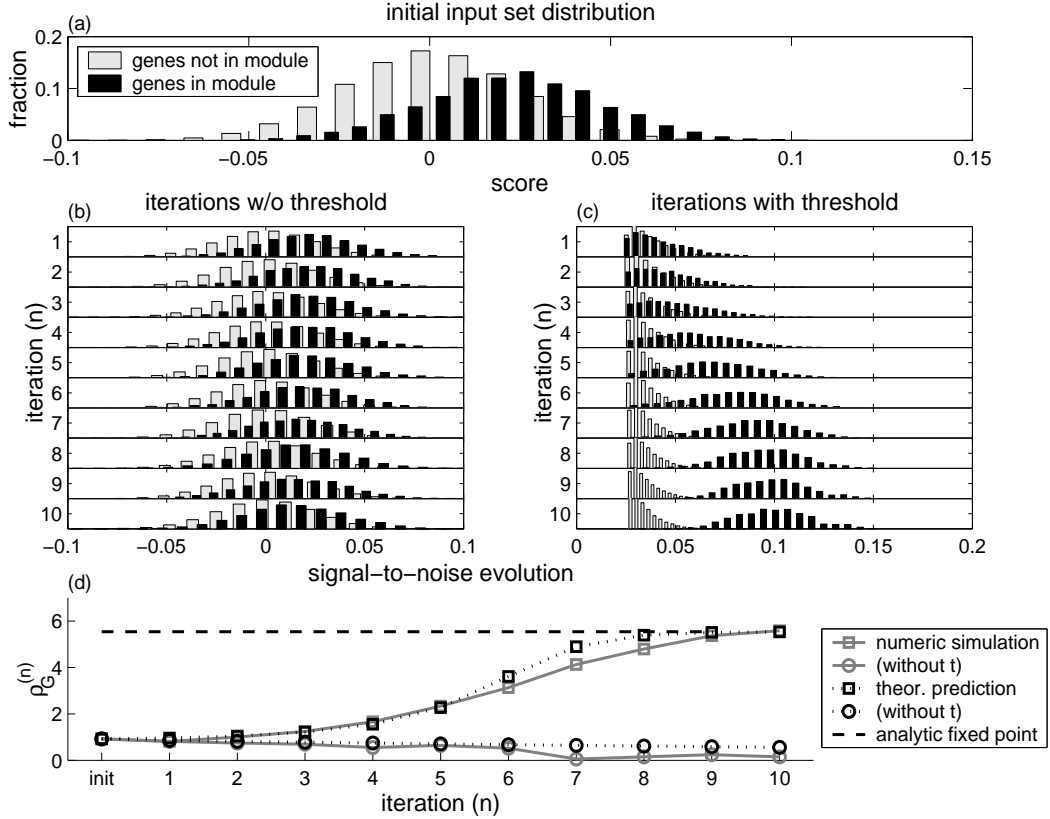,angle=0,width=14cm,height=11cm}}
\end{center}
\vspace{1cm} 
\caption{{\it Evolution of the score distributions under the ISA}. (a)~The distributions of the gene scores of 100 input sets which serve as seeds for the iterations of our algorithm: The distribution of the genes that are not part of the TM (light gray) has a vanishing mean value. The genes belonging to the module (black) are distributed with a positive mean value. Note that the two initial distributions cannot be distinguished from each other accurately. (b$-$c)~Evolution of the two distributions under the iterative scheme defined by eq.~(\ref{iterate-g}). (b) Without applying a threshold, the mean of the signal-distribution decreases in each iteration and the separability of the two distributions does not improve. (c)~When a threshold ($t=1$) is applied the mean of the signal distribution increases in each step until it saturates at a value where the two distributions are well separated. (d)~The signal-to-noise ratio $\rho_G^{(n)}$ characterizes the separability between the gene score distributions for the genes within and outside the module (c.f. section~\ref{analysis} for details). We plot $\rho_G^{(n)}$ as a function of the number of iterations $n$. The evolution of $\rho_G^{(n)}$ under the iterations scheme with (squares) and without (circles) a threshold obtained from the numerical simulation (gray) are in good agreement with the theoretical predictions (black) according to eq.~(\ref{approximate-recursions}). We used $N_G=1700$, $N_G^{(m)}=40$ and $\rho_{\cal C}=1$ for this figure.}
\label{evolution}
\end{figure}

We have also performed numerical simulations of the iterative scheme in eq.~(\ref{iterate-g-threshold}). To this end we
employed {\it in-silico} expression data that were generated according to eq.~(\ref{mean-C}) and superimposed with a certain level of noise. The initial gene sets were composed such that only the distribution of the genes scores associated with the module had a non-zero mean value, while the distribution of the remaining genes was centered 
around zero. The simulation allowed us to trace the evolution of the two distributions under the iterations. The results indicate a good agreement between the numerical and the analytical results. Details of this analysis are presented in Fig.~\ref{evolution}. In particular, in Fig.~\ref{evolution}d we show an example where only the application of a proper threshold leads to a separation between the two distributions.

%=================================
\section{Beyond the single module}
\label{beyond-single-module}
%=================================

In order to study the ISA in a more realistic scenario, we have performed further numerical simulations
based on {\it in-silico} expression data encoding several, possibly overlapping transcription modules.
These data were generated according to the following simple model: Each module $M_m$ is governed by a single (virtual) transcription factor whose activity is described by a pair of vectors $\{\bg_m,\bc_m\}$. The non-zero components $g_m^{(g)}$ of the gene-vector $\bg_m$ specify the genes that are transcribed if the transcription factor $m$ is active, while the non-zero components $c_m^{(c)}$ of the condition-vector $\bc_m$ specify the conditions that activate this transcription factor. Then for $N_M$ modules the log expression of gene $g$ at condition $c$ is defined as $E^{cg} = \sum_{m=1}^{N_M} g_m^{(g)} c_m^{(c)}$. The final expression matrix is obtained by adding noise to these matrix elements.

%--------------------------------------------------------
\subsection{Expression data corresponding to two modules}
%--------------------------------------------------------

\begin{figure}[\figpos]
\begin{center}
  \mbox{\epsfig{figure=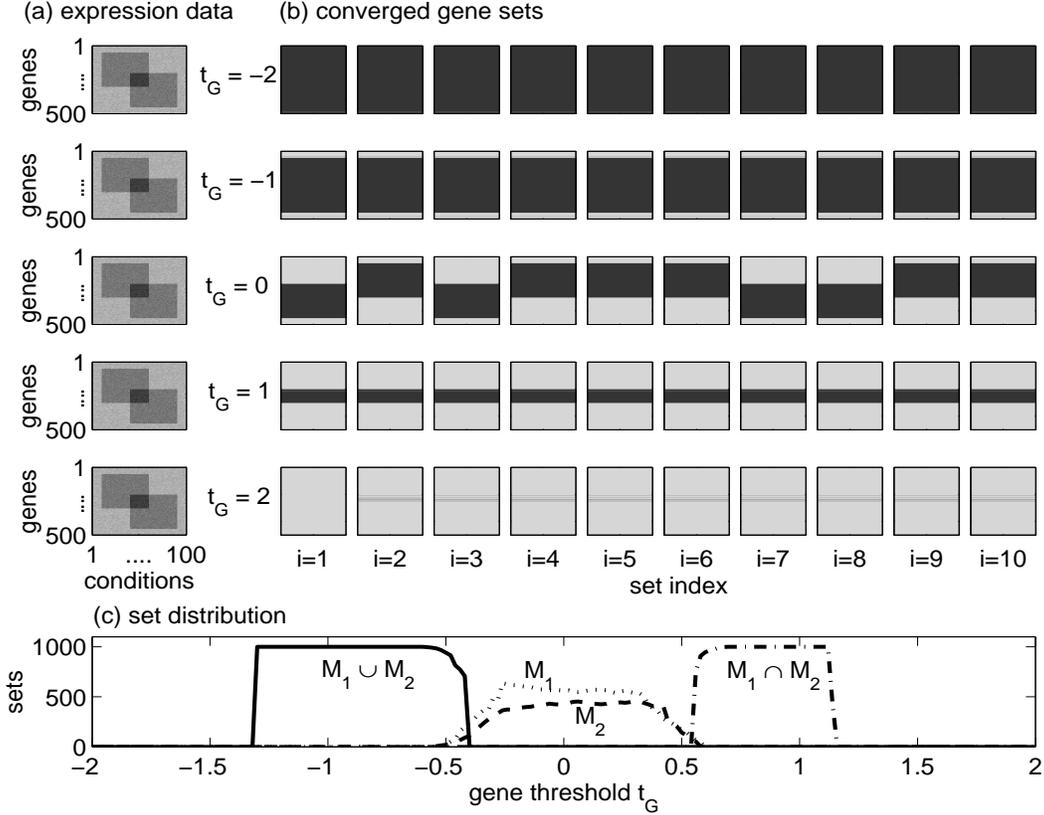,angle=0,width=14cm,height=11.0cm}}
\end{center}
\vspace{0.0cm} 
\caption{{\it Identification of overlapping modules}. An {\it in-silico} expression matrix describing 500 genes under 100 experimental conditions was generated according to the model introduced in the text. 
The data corresponds to two overlapping transcription modules $M_1$ and $M_2$, each containing 250 genes and 50 conditions. (a)~The expression matrix is shown for comparison on the left of each row. (b$-$c)~Using this matrix we applied the ISA to 1000 input sets composed of randomly chosen genes. Iterations were performed using different choices of the threshold $t_G$. (b)~The boxes in each row represent 10 of the resulting converged gene sets, that were obtained for $t_G$ as indicated on the left. Each box $i=1,...,10$ is composed of 500 lines that specify the genes which appear in the corresponding fixed point. Genes that belong to the converged set are represented by a dark gray line, while the remaining genes are shown in light gray. For $t_G \simeq -2$ the output sets contain all the genes, $t_G \simeq -1$ yields output sets containing the genes that are associated with either of the two modules, for $t_G \simeq 0$ there are two types of output sets, comprising either the genes of $M_1$ or of $M_2$, for $t_G \simeq 1$ all the output set contain only those genes that belong to both modules and for $t_G \simeq 2$ the output sets are essentially empty. (c)~The number of sets that converged (within 95\% accuracy) to $M_1 \bigcup M_1$ (solid), $M_1$ (dotted), $M_2$ (dashed) or $M_1 \bigcap M_1$ (dash-dotted) are plotted as a function of $t_G$. Scanning over different thresholds reveals the modular structure of the expression data ($M_1 \bigcup M_1 \to M_1, M_2 \to M_1 \bigcap M_1$). }
\label{two-modules}
\end{figure}

As initial example we consider {\it in-silico} expression data based on two transcription factors. We defined the components $c_m^{(c)}$ and  $g_m^{(g)}$ for $m=1,2$ such that there are two overlapping transcription modules $M_1$ and $M_2$ (see Fig.~\ref{two-modules} for details). We applied the ISA to a collection of input sets composed of randomly chosen genes. We found that the structure of the resulting fixed points depends strongly on the threshold $t_G$. Fig.~\ref{two-modules}b shows the corresponding output sets for a discrete choice thresholds: For a very low threshold ($t \simeq -2$) the output sets contain essentially all the genes. Applying a somewhat higher threshold ($t \simeq -1$) yields output sets containing all the genes that are associated with either of the two modules. For a moderate threshold ($t \simeq 0$) there are two types of output sets, comprising either the genes of $M_1$ or $M_2$. For a high threshold ($t \simeq 1$) all the output set contain only those genes that belong to both modules. Finally, for a very high threshold ($t \simeq 2$) the output sets are empty. For intermediate values of the threshold value one observes relatively sharp transitions between these well-defined fixed points (Fig.~\ref{two-modules}c). At these transitions the correspondence between the output sets and the modular structure of the data is less precise.

We have also varied the condition threshold $t_C$. Interestingly, for not too large a threshold ($t_C \lsim 2$) the resulting gene output sets are almost independent of the choice of $t_C$. However, the condition output sets depend critically on the value of $t_C$ and exhibit a similar behavior as the gene output sets in terms of structure (not shown). This is not surprising, since the ISA is symmetric with respect to genes and conditions. We conclude that scanning over different values of $t_G$ and $t_C$ reveals the modular structure of the expression data, starting from the ``supermodule'' $M_1 \bigcup M_1$, over its overlapping components $M_1$ and $M_2$, to the ``submodule'' $M_1 \bigcap M_1$.

%---------------------------------------------------------
\subsection{Expression data corresponding to many modules}
%---------------------------------------------------------

The above example shows that the ISA can identify overlapping modules. However, for $N_M=2$ there exist
only $2^2=4$ possible transcriptional states, so the 100 conditions of the expression data are highly redundant.
For real data the situation is reverse: The number of experimental conditions is much smaller than the possible
number of transcriptional states. In order to study how the ISA deals with such a scenario we considered a set of more realistic models based on many transcription modules. We investigated to what extend the
ISA, as well as hierarchical clustering and SVD, were able to reconstruct these modules from the respective
{\it in-silico} expression data.

In the first numerical experiment we studied how the different algorithms handle noisy data. To this end we generated expression matrices corresponding to 1050 genes and 1000 experimental conditions that belong to 25 modules of different sizes, each associated with a transcription factor. In order to focus on the effect of noise we considered only non-overlapping modules that do not share any genes or conditions. Onto the binary expression data we superimposed noise from a random distribution. We varied the width $\sigma$ of this distribution, simulating different levels of noise. 

\begin{figure}[\figpos]
\begin{center}
  \mbox{\epsfig{figure=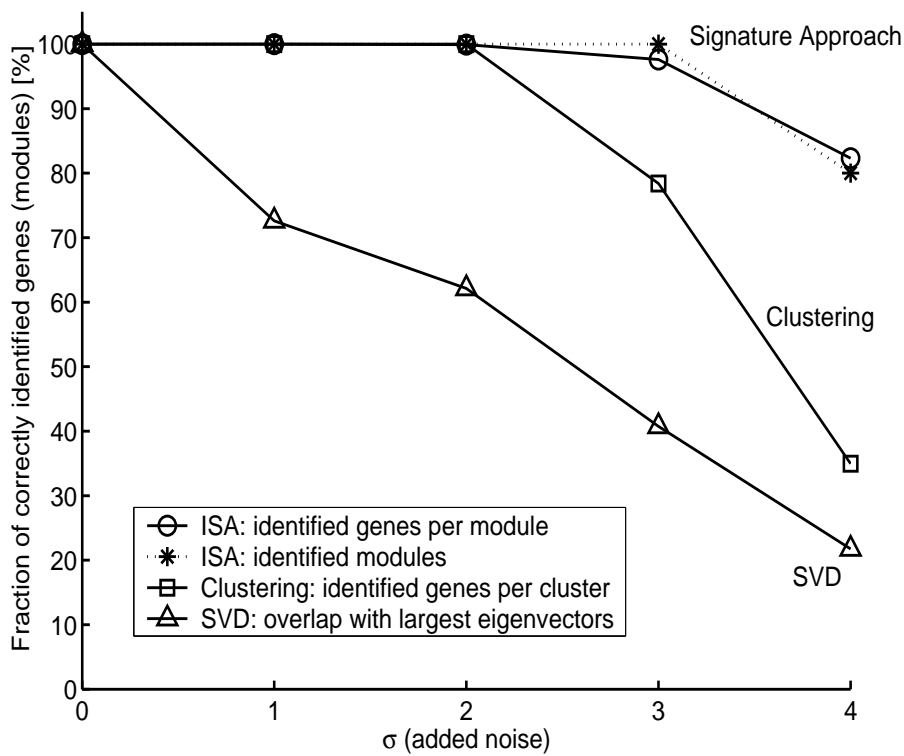,angle=0,width=12cm,height=10cm}}
\end{center}
\vspace{1cm} 
\caption{{\it Module identification from noisy expression data}. {\it In-silico} expression matrices for 1050 genes under 1000 conditions, corresponding to 25 non-overlapping transcription modules of different sizes, were generated according to the model described in the text. Noise from a uniform distribution was superimposed onto this expression data. The width $\sigma$ of this noise distribution was varied, simulating different levels of noise. We quantified the efficiency of different algorithms to retrieve the modules from the expression data as described in the text. We show the fraction of correctly identified genes for the ISA (circles), hierarchical clustering (squares) and SVD (triangles). For the ISA we also the fraction of correctly identified modules are indicated (asterisks). SVD is very sensitive to the addition of noise and fails to identify the modules accurately, even for a small level of noise. Clustering can handle a moderate amount of noise, but not as much as the ISA.} 
\label{noise}
\end{figure}

In order to quantify how well the modules were identified by the different methods we proceeded as follows: For SVD we collected the 25 eigenvectors of the gene-gene correlation matrix that were associated with the largest eigenvalues. For each of the 25 modules we selected the eigenvector that had the largest overlap with the gene-vector characterizing the module, and in Fig.~\ref{noise} we show the average Pearson coefficient between these two vectors (triangles). For hierarchical clustering we used the matlab implementation for average linkage to compute the complete hierarchical cluster tree. Using this cluster tree we partitioned the expression matrix using different cutoffs such that the resultant partitions contained at least 15 and at most 40 clusters. From all these partitions we selected the one whose clusters had the highest average overlap with the gene content of the modules. This overlap is shown in Fig.~\ref{noise} (squares). Finally, for the ISA we re-constructed the modules from the fixed points that occurred repeatedly. Namely, in order to avoid artifacts due to distinct, but very similar fixed points, we ``fused'' these solution using a procedure that resembles agglomerative clustering, albeit for modules rather than genes (see Ref.~\cite{Ihmels2002} for details). The fraction of correctly identified genes per module (circles) as well as the fraction of correctly identified modules (asterisks) is shown in Fig.~\ref{noise}. We conclude that for noisy data the identification capability of the ISA is superior to that of SVD and clustering. In particular, SVD is very sensitive to the addition of noise and fails to identify the modules accurately, even for a small level of noise. Clustering can handle a moderate amount of noise, but not as much as the ISA.

\begin{figure}[\figpos]
\begin{center}
  \mbox{\epsfig{figure=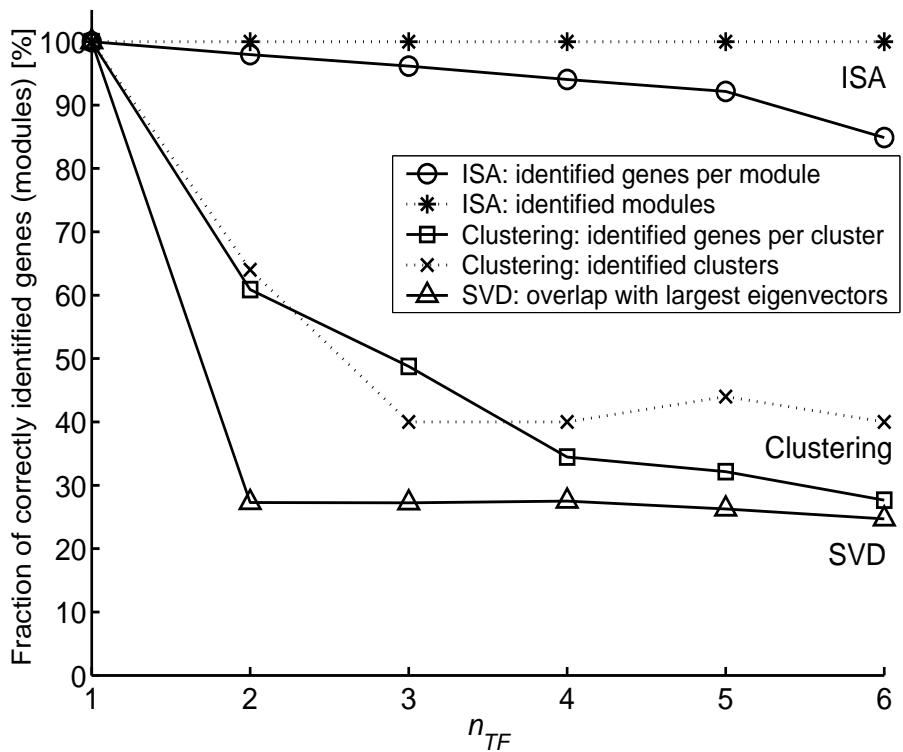,angle=0,width=12cm,height=10cm}}
\end{center}
\vspace{0.5cm} \caption{{\it Module identification in the presence of combinatorial regulation}. 
{\it In-silico} expression matrices corresponding to 25 overlapping modules were generated according to a model that allows for combinatorial regulation (see text for details). The degree of overlap between the modules is specified by the average number of transcription factors involved in the regulation of each gene ($n_{TF}$). Only for $n_{TF} = 1$ each gene is associated with exactly one transcription factor. For larger values of $n_{TF}$ distinct modules share common genes. We applied the SVD, hierarchical clustering (see Ref.~\cite{Ihmels2002} for related results) and the ISA to the expression matrices generated for $n_{TF} = 1, ..., 6$ and evaluated the outputs as described in the text. The ISA could successfully identify all the transcription modules even in the case of highly overlapping modules (asterisks), The fraction of correctly identified genes per module only decreases slightly as a function of $n_{TF}$ (circles). In contrast, for $n_{TF} > 1$ the identification capabilities of clustering (squares/crosses) and SVD (triangles) rapidly decrease. This is because the clustering algorithm does not allow for multiple assignments of one gene to different modules and therefore usually captures only small, incomplete fractions of the overlapping modules. Similarly, if the expression matrix cannot be reorganized into block-diagonal shape due to the overlap between the modules, the eigenvectors identified by SVD fail to characterize the modules properly.} 
\label{complexity}
\end{figure}

A second numerical experiment was designed to study quantitatively the ability to identify overlapping modules. We specify the regulatory complexity by the the number of transcription factors per gene $n_{TF}$. Only if each gene (and condition) is associated with exactly one transcription factor ($n_{TF} = 1$) the expression matrix can be written in block-diagonal form. For larger values of $n_{TF}$ distinct modules share common genes and conditions and the expression matrix cannot be reorganized into in block-diagonal shape. We applied the SVD, hierarchical clustering and the ISA to the expression matrices generated for $n_{TF} = 1, ..., 6$ and evaluated the outputs in the same manner as described above (see Ref.~\cite{Ihmels2002} for related results). The results are shown in Fig.~\ref{complexity}. One can see that the ISA could successfully identify all the transcription modules even in the case of highly overlapping modules. In contrast, for $n_{TF} > 1$ the identification capabilities of SVD and clustering rapidly decrease. This is because the clustering algorithm does not allow for multiple assignments of one gene to different modules and therefore usually captures only small, incomplete fractions of the overlapping modules. Similarly, if the expression matrix cannot be reorganized into block-diagonal shape due to the overlap between the modules, the eigenvectors identified by SVD fail to characterize the modules properly.

%==================================================
\section{Applying the ISA to yeast expression data}
\label{application}
%==================================================

The analytical and numerical studies presented above indicate that the ISA is well-suited for the analysis of expression data. In this section we give a brief presentation of the biological insight that can be obtained from applying our method to real data. We analyzed a diverse set of more than 1000 DNA-chip experiments that were obtained by different groups~\cite{yeast}. The yeast S. cerevisiae is an ideal model organism to test our algorithm, due to the wealth of expression data and the large amount additional biological knowledge that exists for this organism.

We have applied the ISA  to the yeast expression data using different values for the gene-threshold $t_G = 1.8, 1.9, ..., 4.0$, while the condition-threshold was fixed to $t_C = 2.0$. (As we pointed out previously the gene-content of the modules depends only weakly on the exact choice for $t_C$.) For each value of $t_G$ we employed $\sim 20,000$ randomly composed initial gene sets of various sizes in the search for fixed points.  The modules were reconstructed from the recurrent fixed points using a similar algorithm as for the {\it in-silico} expression data. Indeed such a processing of the ``raw'' fixed points is needed to avoid many similar modules that biologically correspond to the same co-regulated unit.

The number of modules increases with $t_G$, ranging between five at the lowest level ($t_G=1.8$) to $\sim 100$ at the highest resolution ($t_G=4$). In contrast, the typical module size declines rapidly as a function of $t_G$. The step-wise increasing of $t_G$ exposed many chains of closely related modules that persist for finite ranges $t_G \in [t_G^{bottom},t_G^{top}]$. Increasing $t_G$, the number of genes assigned to each element of the chain decreases until the size of the module declines sharply at $t_G = t_G^{top}$ and either disappears completely or splits into two or more sub-modules. Likewise decreasing $t_G$ beyond $t_G^{bottom}$ destabilizes the fixed point, since many unrelated genes are added to the module that pull the module towards a different fixed point. In this case the module may either `merge' with another module or flow into a completely different fixed point. 

The five stable fixed points identified for $t_G = 1.8$ correspond to the central functions of the yeast organism: protein synthesis, cell-cycle (G1), mating, amino-acid biosynthesis and stress response. Each module contains between 100 and 300 genes. Protein synthesis and stress are the most dominant modules and comprise most of the experimental conditions of the data set. In fact, these modules remain fixed points throughout the entire range of thresholds considered here, and therefore can be considered the backbone of the transcriptional network. 

\begin{figure}[\figpos]
\begin{center}
  \mbox{\epsfig{figure=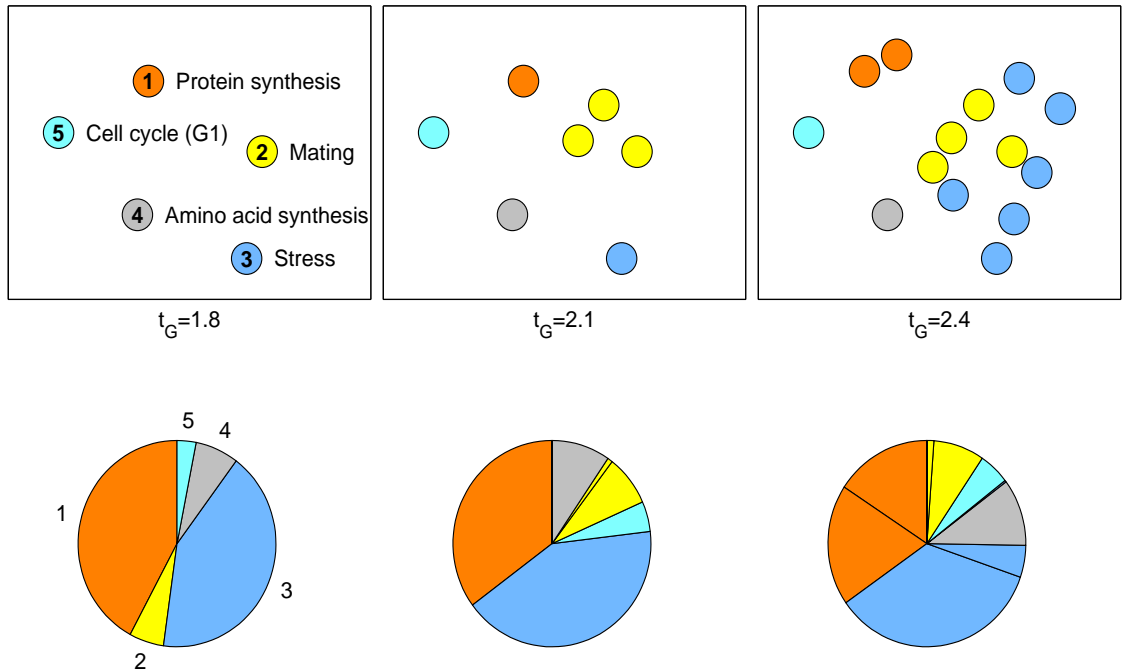,angle=0,width=15cm,height=9cm}}
\end{center}
\vspace{1cm} \caption{{\it Modular organization of yeast expression data}. The iterative signature algorithm was applied to genome wide yeast expression data gathered by more than 1000 DNA-chip experiments. (a) The figure shows the identified modules at three different gene-thresholds $t_G=\{1.8, 2.1, 2.4 \}$. 
For each threshold the corresponding modules are displayed in a plane, such that their distance reflects their correlation with respect to conditions. Moving to a higher threshold, corresponding of modules are kept in the same position in each plane, while the ``new'' modules are placed such that their position reflects best their correlation with the other modules. The left-most plane corresponds to the lowest threshold ($t_G=1.8$), where only five fixed points exist. The corresponding modules can be associated with central functions of the yeast organism: protein synthesis, cell-cycle (G1), mating, amino-acid biosynthesis and stress response. We use color coding to indicate which of the fixed points that emerge at higher thresholds are related to these five central modules (i.e. they would convergence to the respective module at the lowest threshold). b) The pie charts show for the number of random input sets that converged to the respective fixed point. The color coding is as in (a).}
\label{yeast}
\end{figure}

A visualization of this network is presented in Fig.~\ref{yeast}a. For each threshold the corresponding modules are displayed in a plane, such that their distance reflects their correlation with respect to conditions. Moving to a higher threshold, nested sets of modules are kept in the same position in each plane, while the ``new'' modules are placed such that their position reflects best their correlation with the other modules. This organization of the chains of nested modules is somewhat similar to the data presentation by hierarchical trees commonly produced by cluster algorithms. However, in our case, chains of modules may extend over a finite range of $t_G$ and distinct chains can contain common genes. Additional information, such as the number of input seeds that converged to the same fixed pointed (shown as pie charts in Fig.~\ref{yeast}b), provide further inside into the transcriptional network.

In a previous analysis of the same data~\cite{Ihmels2002} we applied the map in eqs.~(\ref{cond-signature}) and~(\ref{gene-signature}) to a variety of biologically motivated input-sets $\{\bg^{(0)}_i\}$ assembled according to prior knowledge of the regulatory sequence or function of the genes, and reconstructed the modules from recurrent realizations of the output-sets defined by $\bg^{(1)}$ and $\bc^{(1)}$. Remarkably, the ISA (which requires no information beyond the expression data whatsoever) revealed essentially all the co-regulated units that we found in this analysis, as well as several new transcription modules that had not been identified previously. Moreover, the ISA provides additional insight into the modular organization through the evolution of the modules over different threshold values. Studying the functional annotations of the genes assigned to the modules, we observed a strong coherence for the genes that have been annotated in most of these modules. This suggests that the ISA provides a biologically meaningful decomposition into co-regulated units. A comprehensive discussion of the biological implications of this analysis is beyond the scope of this work and will be pursued elsewhere~\cite{Bergmann}.

%====================
\section{Conclusions}
%====================
\label{conclusions}

We have presented a novel method for the analysis of gene expression data. The innovation of our approach is twofold: On the conceptual level we provide a rigorous definition of what we want to extract from the expression data by introducing the notion of a {\it transcription module} (TM). Our definition in eq.~(\ref{consistency}) assigns to a TM both a set of co-regulated genes and the set of experimental conditions under which this co-regulation is the most stringent. The size of a TM depends critically on the associated set of two thresholds that determine the similarity between the genes and conditions of the module, respectively. The genes and conditions of a TM are mutually consistent implying that the latter can be obtained from the former and vice versa. The notion of a TM is well motivated biologically. Ideally the genes and conditions can be associated with a transcription factor or a (fraction of) a pathway. Importantly distinct modules may share both common genes and conditions. 

On the computational level our definition of a TM provides the basis for simple, but efficient algorithm to obtain the modules encoded in the expression data. Starting from a set of randomly selected genes (or conditions) one refines iteratively the genes and conditions until they are mutually consistent and match the definition of a TM. The important point is that at each step of the iterations we apply a threshold function, thus maintaining only significantly co-regulated genes and the associated co-regulating conditions. The threshold stabilizes compact sets of co-regulated genes and prevents the introduction of noise from unrelated genes and conditions. Using a sufficiently large number of initial random sets it is possible to determine all the fixed points of the iterative scheme for a given pair of thresholds. Scanning through a range of values for these thresholds decomposes the data into modules at different resolutions. Since the computation time for each iteration of our algorithm scales only linearly with the total number of genes it is particularly well-suited for the analysis of large scale expression data.

Considering a simplified scenario of a single transcription module embedded in a noisy background of unrelated genes, we showed analytically that the application of a threshold improves the convergence properties of the iterative scheme. Specifically, we considered the gene-vector that undergoes iterations as a stochastic entity and studied the evolution of its distribution under the iterations for a given threshold. This allowed us to quantify how the successful identification of the module depends on the size of the module and the noise in the expression data. 

Our analytical insights were confirmed numerically using computer-generated expression data. More complex gene regulation were also simulated {\it in-silico}. Considering a model with two overlapping transcription modules, we showed that applying the ISA using a range of threshold values reveals the structure of the expression data at different resolutions. Depending on the value of the threshold our algorithm can reveal each of the two modules, as well as their union and intersection. Using large computer-generated expression matrices we studied the capability of the ISA to reveal a large number of overlapping transcription modules from noisy expression data. We find that our method is significantly more efficient at this task than standard tools, like SVD and clustering.

The threshold functions as a resolution parameter in our analysis of real expression data. Using genome-wide expression data gathered in more than 1000 experimental conditions, we decomposed the yeast genome into sets of transcription modules at different resolutions. The modular decomposition reveals a hierarchical structure of the regulatory network. At the lowest resolution we identified five transcription modules that correspond to the central functions of the yeast organism. Increasing the threshold the number of modules increases while their size decreases. The functional coherence of these modules indicates both the reliability of our approach and the strong correlation between co-function and co-regulation at the transcriptional level in yeast. A comprehensive discussion of the biological implications of this analysis will be presented elsewhere~\cite{Bergmann}.

Finally we note that our formalism can be applied to analyze any data set that consists of multi-component measurements. While we presented our method in the context of gene-expression data, it is clear that our approach is well-suited to reveal the modular organization encoded in any data matrix. Applications of the ISA could include the analysis of biological data on protein-protein interactions or cell growth assays, as well as other large scale data, where a meaningful reduction of complexity is needed.

%%%%%%%%%%%%%%%%%
%\acknowledgements
%%%%%%%%%%%%%%%%%

\vspace{5mm}
\noindent 
{\bf Acknowledgements}:
We thank J. Doyle for bringing our attention to the similarity between SVD and the ISA. We thank E. Domany, Y. Kafri and S. Shnider for discussions and comments on the manuscript. This work was supported by the NIH grant \#A150562, the Israeli Science Ministry and the Benoziyo center. S.\,B. is a Koshland fellow. N.\,B. is the incumbent of the Soretta and Henry Shapiro career development chair.

%%%%%%%%%
\appendix
\section{Appendix}
%%%%%%%%%

%----------------------------------------
\subsection{Singular Value Decomposition}
%----------------------------------------
\label{SVD-appendix}

This appendix reviews Singular Value Decomposition (SVD), which is a common tool for the analysis of expression 
data. We use notations that make the similarities with the Iterative Signature Algorithm (ISA) the most apparent.
SVD is used to reduce the dimensionality of the data by projecting it onto a subspace in such a way that as little information is lost as possible. To this end consider the following matrix:
\beq \label{E_m}
\bE_m = \bc_m \, \bg_m^T \,,
\eeq
whose elements $E_m^{cg} = g_m^{(g)} \, c_m^{(c)}$ are simply the products of the components of a given gene-vector $\bg_m$ and condition-vectors $\bc_m$. For two binary vectors $\bg_m$ and $\bc_m$ (whose elements are either zero or one) $E_m^{cg}$ is unity if the module $m$ contains the gene $g$ and the condition $c$ (i.e. the relevant vector components are $g_m^{(g)}=1$ and $c_m^{(c)}=1$). For real vectors $\bg_m \in \IR^{N_G}$ and $\bc_m \in \IR^{N_C}$ it is useful to rewrite the matrix in eq.~(\ref{E_m}) as
\beq 
\bE_m = \mu_m \hc_m \, \hg_m^T \,,
\eeq
in terms of the normalized vectors $\hg_m = \bg_m/|\bg_m|$ and $\hc_m = \bc_m /|\bc_m|$. This normalization removes the ambiguity in the choice of $\bg_m$ and $\bc_m$ due to the invariance of $\bE_m$ under the transformation $\bg_m \to \phi \, \bg_m$ and $\bc_m \to \bc_m / \phi$, where $\phi \ne 0$ is an arbitrary real number. The prefactor $\mu_m = |\bg_m| \, |\bc_m|$ is just the product of the lengths of $\bg_m$ and $\bc_m$. Then each module is associated with a triple $(\mu_m, \hg_m, \hc_m)$ of a real number and two normalized vectors. Comparing the magnitude of any two matrix elements $E_m^{cg}$ and $E_m^{g'c'}$ reveals the relative importance between the gene-condition pairs $(g,c)$ and $(g',c')$ for module $m$.

Multiplying $\bE_m$ with an arbitrary gene-vector $\bg$ gives
\beq
\bE_m \, \bg = \alpha \, \hc_m ~~~\mbox{with}~~~ \alpha = \mu_m \, \hg_m^T \, \bg \,,
\eeq
while multiplication of $\bE_m^T = \mu_m \, \hg_m \, \hc_m^T $ with any condition-vector $\bc$ gives
\beq
\bE_m^T \, \bc = \beta \, \hg_m ~~~\mbox{with}~~~ \beta = \mu_m \, \hc_m^T \, \bc \,.
\eeq
Thus $\bE_m$ and $\bE_m^T$ are projection operators onto the one-dimensional spaces spanned by $\hg_m$ and $\hc_m$, respectively. Consequently theses matrices have rank 1. 

Now the basic idea of SVD is to reduce the complexity of the data by expressing $\bE$ in terms of a relatively small number $N_M (\ll N_G, N_C)$ of such rank 1 matrices:
\beq \label{expand-E}
\bE = \sum_m^{N_M} \bE_m + \bR_{N_M} \,.
\eeq
Here $\bR$ denotes the residual term whose euklidean norm $|\bR| = \sqrt{\sum_{g,c} (R^{cg})^2}$ has to be minimized in order to optimize the decomposition into modules in the above equation.

It is instructive to consider first the minimization for the case $N_M=1$. We have
\beqs \label{R}
|\bR|^2 &=& \sum_{g,c} (E^{cg}-E_m^{cg})^2 = 
\sum_{g,c} (E^{cg}-\mu_m \hat c_m^{(c)} \hat g_m^{(g)})^2 \\
&=& \sum_{g,c} (E^{cg})^2-2\mu_m E^{cg} \hat c_m^{(c)} \hat g_m^{(g)} +  
\mu_m^2 (\hat c_m^{(c)} \hat g_m^{(g)})^2 \,.
\eeqs
Setting the derivative of $|\bR|^2$ with respect to the component $\hat c_m^{(c)}$,
\beq
{\partial |\bR|^2 \over \hat c_m^{(c)}} =
\sum_g -2\mu_m E^{cg} g_m^{(g)} + 2\mu_m^2 (\hat g_m^{(g)})^2 \hat c_m^{(c)} \,,
\eeq
to zero we find that that $\mu_m \hat c_m^{(c)} = \sum_g E^{cg} g_m^{(g)} / \sum_g (g_m^{(g)})^2$
or, recalling the normalization of $\hg_m$ and switching to vector notation:
\beq \label{eigen-c}
\mu_m \hc_m = \bE \, \hg_m \,.
\eeq
Similarly equating ${\partial |\bR|^2 / \hat g_m^{(g)}}$ to zero it follows that
\beq \label{eigen-g}
\mu_m \hg_m = \bE^T \hc_m \,.
\eeq
This remarkable result implies that $\bE_m$ can be determined simply by solving simultaneously the linear equations in eqs.~(\ref{eigen-c}) and~(\ref{eigen-g}). The latter is equivalent to a singular value decomposition (SVD) of the matrix $\bE$:
\beq
\bG^T \bE \bC = \bM \,,
\eeq
where $\bG = (\hg_1, \hg_2, ..., \hg_r)$ and $\bC = (\hc_1, \hc_2, ..., \hc_r)$ are orthogonal matrices. $\bM$ is a diagonal matrix of the same dimensions as $\bE$ whose non-zero elements are given by $\mu_m$ and ordered such that $\mu_1^2 \ge \mu_2^2 \ge ... \ge \mu_r^2$. $r \le \min(N_G,N_C)$ is the rank of the expression matrix $\bE$. Combining eqs.~(\ref{eigen-c}) and~(\ref{eigen-g}) one finds
\beqs 
\label{eigen-genes}
\bE^T \bE \hg_m = \mu^2 \hg_m \,, \\
\bE \bE^T \hc_m = \mu^2 \hc_m \,,
\eeqs
implying that $\bG$ is composed of the eigenvectors $\hg_m$ of $\bE^T \bE$ and $\bC$ consist of the eigenvectors $\hc_m$ of $\bE \bE^T$. One way to solve the above equations is start with some initial gene-vector $\hg^{(0)}$, obtain the corresponding condition-vector via $\hc^{(1)} = \bE \hg^{(0)} / |\bE \hg^{(0)}|$ according to eq.~(\ref{eigen-c}), and use the result to compute $\hg^{(1)} = \bE^T \hc^{(1)} / |\bE^T \hc^{(1)}|$ using eq.~(\ref{eigen-g}). Iterating this alternating procedure as in eqs.~(\ref{find-c}) and~(\ref{find-g}) converges to the pair $(\hg_1,\hc_1)$ associated with largest eigenvalue $\mu_1^2 = |\bE \hg_1|^2$ provided that the initial vector $\hg^{(0)}$ was not orthogonal to $\hg_1$. Thus the predominant module emerges as the ``fixed point'' of the above coupled equations.

From eq.~(\ref{R}) it follows that $|\bR|^2 = \sum_{g,c} (E^{cg})^2 - \mu_m^2$. Hence for $N_M=1$ the norm of the residual term, $|\bR|^2$, is minimized exactly by the triple $(\mu_1, \hg_1,\hc_1)$. It is straightforward to extend this approach to the expansion of the expression matrix in terms of several modules as in eq.~(\ref{expand-E}). To this end one first computes $\bE_1 = \mu_1 \hc_1 \hg_1^T$ as described above and applies the same scheme to the residual term $\bR_1 = \bE - \bE_1$. This yields $\bE_2 = \mu_2 \hc_2 \hg_2^T$  associated with the second largest eigenvalue $\mu_2$. Repeating this procedure {\it sequentially} yields eventually the complete SVD of the matrix $\bE$. However, for practical purposes it is usually sufficient to compute only a limited numbers of triples $(\mu_m,\hg_m,\hc_m)$ with $m=1, ..., N_M$ until the norm of the residual term $|\bR_{N_M}|^2 = \sum_{g,c} (E^{cg})^2 - \sum_{m=1}^{N_M} \mu_m^2$ is below a certain threshold. Thus, approximating the expression matrix in terms of a relatively small number of modules, $N_M \ll r$ reduces the complexity of the data. 

There are two interpretations for the expansion in eq.~(\ref{expand-E}) that depend on the way the 
expression data is viewed. If we consider the data as a collection of gene-vectors $\bg_c$ as in eq.~(\ref{E-g}),
then eq.~(\ref{expand-E}) translates into an expansion of these vectors in terms of a collection of gene-vectors, i.e. 
\beq
\bg_c = \sum_{m=1}^{N_M} \mu_m \hat c^{(c)}_m \hg_m + \bg_c^R ~~~~(c=1,...,N_C) \,,
\eeq
where $\{\hg_m\}$ is the basis (one for {\it all} $\bg_c$), and the expansion coefficients are given by $\mu_m \hat c^{(c)}_m$ (one for {\it each} $\bg_c$). Moreover, for each $\bg_c$ there is a residual gene-vector $\bg_c^R$, that determines how well $\bg_c$ is approximated by the sum. Conversely, if we consider the data as a collection of condition-vectors $\bc_g$ as in eq.~(\ref{E-c}), then the expansion in eq.~(\ref{expand-E}) can be read as
\beq
\bc_g = \sum_{m=1}^{N_M} \mu_m \hat g^{(g)}_m \hc_m + \bc_g^R ~~~~(g=1,...,N_G) \,,
\eeq
where $\bc_g^R$ denotes the residual condition-vector. In this case the condition-vectors of the modules, $\{\hc_m\}$, provide the basis of expansion, while the expansion coefficients for each $\bc_g$ are given by $\mu_m \hat g^{(g)}_m$. 

So far we have left the normalization of $\bE$ unspecified. In fact the choice of normalization follows from the interpretation of the data, if, instead of a minimal residual term in eq.~(\ref{R}), one demands maximal variance among the {\it principal components} (the projections of the data rows or columns onto the eigenvectors associated with the largest eigenvalues). For example, if the expression data is viewed as a collection of gene-vectors, one would like to find the vector $\hg_1$ that maximizes the variance of the principal components $c^{(c)}_1 = \bg_c^T \hg_1$, i.e.
\beq
V_1^g = {1 \over N_C} \sum_{c=1}^{N_C} \left(c^{(c)}_1 - \bigl<c^{(c)}_1\bigr>_c \right)^2 
= {1 \over N_C} \hg_1^T \bS_g \hg_1 \,.
\eeq
Here the bilinear term has been written in terms of the scatter matrix
\beq
\bS_g \equiv \sum_{c=1}^{N_C} \left(\bg_c - \bigl<\bg_c\bigr>_c \right) \,
\left(\bg_c - \bigl<\bg_c\bigr>_c \right)^T \,.
\eeq
Maximizing $V_1^g$ under the constraint that $\hg_1^T \hg_1 = 1$ is equivalent
to finding the eigenvector of $\bS_g$ associated with the largest eigenvalue.
For normalized data, $\bS_g$ coincides with the gene-gene correlation matrix
\beq
\cC_g = \bE_C^T \, \bE_C ~~~\mbox{with}~~~ {\cal C}_g^{gg'} = \hc_g^T \hc_{g'} \,.
\eeq
Conversely, if the expression data is viewed as a collection of condition-vectors, 
the vector $\hc_1$ that maximizes the variance of the components $g^{(g)}_1 = \bc_g^T \hc_1$,
is the eigenvector associated with the largest eigenvalue of the scatter matrix
\beq
\bS_c \equiv \sum_{g=1}^{N_G} \left(\bc_g - \bigl<\bc_g\bigr>_g \right) \, 
\left(\bc_g - \bigl<\bc_g\bigr>_g \right)^T \,.
\eeq
For normalized data, $\bS_c$ equals to the condition-condition correlation matrix
\beq
\cC_c = \bE_G \, \bE_G^T ~~~\mbox{with}~~~ {\cal C}_c^{cc'} = \hg_c^T \hg_{c'} \,.
\eeq
Note, however, that since $\bE_G \ne \bE_C$, the matrices $\bE_C \, \bE_C^T$ and $\bE_G^T \, \bE_G$ 
are different from $\cC_c$ and $\cC_g$, and do not represent correlation matrices.

%%%%%%%%%%%%%%%%%%%%%%%%%%%%%%%%%%%%%%%%%%%%%%%%%%%%%%%%%%
\subsection{The variance of a product of random variables}
\label{variance-product}
%%%%%%%%%%%%%%%%%%%%%%%%%%%%%%%%%%%%%%%%%%%%%%%%%%%%%%%%%%

By definition the mean of the product of two {\it independent} random variables $a$ and $b$ is the product of their mean values, i.e.
\beq \label{multi-mean}
\bigl<a b\bigr> = \bigl<a\bigr> \bigl<b\bigr> \,.
\eeq
Since the expression for the variance of the product $a b$ in eq.~(\ref{multi-mean-var}) may
be somewhat less obvious, we give its derivation here. From the definition of the variance
\beq \label{def-var}
V(a) \equiv \bigl<(a-\bigl<a\bigr>)^2\bigr> = \bigl<a^2\bigr>-\bigl<a\bigr>^2 \,,
\eeq
we obtain
\beqs 
V(a) V(b) &=& \left(\bigl<a^2\bigr>-\bigl<a\bigr>^2\right) \, \left(\bigl<b^2\bigr>-\bigl<b\bigr>^2\right) \\
&=& \bigl<a^2\bigr> \bigl<b^2\bigr> - \bigl<a\bigr>^2 \bigl<b^2\bigr> 
- \bigl<a^2\bigr> \bigl<b\bigr>^2 + \bigl<a\bigr>^2 \bigl<b\bigr>^2 \,.
\label{var-prod}
\eeqs
Then using eqs.~(\ref{multi-mean})-(\ref{var-prod}) it follows that
\beqs
V(a b) &=&  \bigl<a^2 b^2\bigr>-\bigl<a b\bigr>^2 \label{eq1} \\
&=& \bigl<a^2\bigr> \bigl<b^2\bigr> - \bigl<a\bigr>^2 \bigl<b\bigr>^2 \label{eq2} \\
&=& V(a) V(b) + \bigl<a\bigr>^2 \bigl<b^2\bigr> + \bigl<a^2\bigr> \bigl<b\bigr>^2
   -2\bigl<a\bigr>^2 \bigl<b\bigr>^2 \label{eq3} \\
&=& V(a) V(b) 
   +\left(\bigl<a^2\bigr>-\bigl<a\bigr>^2\right) \bigl<b\bigr>^2
   +\left(\bigl<b^2\bigr>-\bigl<b\bigr>^2\right) \bigl<a\bigr>^2 \label{eq4} \\
&=& V(a) V(b) + V(a) \bigl<b\bigr>^2 + V(b) \bigl<a\bigr>^2 \label{eq5} \,.
\eeqs

%%%%%%%%%%%%%%%%%%%%%%%%%%%%%%%%%%%%%%%%%%%%%%%%%%%%%%%
\subsection{Accurate treatment of the noise propagation}
\label{accurate-noise}
%%%%%%%%%%%%%%%%%%%%%%%%%%%%%%%%%%%%%%%%%%%%%%%%%%%%%%%

In order to simplify our presentation of the propagation of the noise under the iterative scheme in eq.~(\ref{iterate-g}) we used the approximate recursive equation in eq.~(\ref{approximate-recursions}) to derive the fixed point noise-to-signal ratio in eq.~(\ref{v-fixed-point}). Here we give an accurate treatment that is valid even if $N_G^{(m)} \gg 1$ is not satisfied.

First, note that if the iterative scheme converges, then for $n \to \infty$ we have $v_G^{(n)} = v_G^{(n-1)} = v_G^{(*)}$ and $\tilde v_G^{(n)} = \tilde v_G^{(n-1)} = \tilde v_G^{(*)}$. In this case we can write two fixed-point equations
\beqs
\label{fix-point-noise}
\tilde v_G^{(*)} \left(1 - {\Delta N_G v_{\cal C} \over (N_G^{(m)})^2}  \right) &=& 
{v_{\cal C} \over N_G^{(m)}} (v_G^{(*)} + 1) \,, \\
\label{fix-point-signal}
v_G^{(*)} \left(1 - {1 \over N_G^{(m)}} \right) &=& \tilde v_G^{(*)}  \,.
\eeqs
Solving eqs.~(\ref{fix-point-noise}) and~(\ref{fix-point-signal}) for $v_G^{(*)}$ we get:
\beq \label{fix-point-solution}
v_G^{(*)} = \left[ \left(1 - {1 \over N_G^{(m)}} \right) \, 
\left({N_G^{(m)} \over v_{\cal C}} - {\Delta N_G \over N_G^{(m)}} \right) - 1 \right]^{-1} 
\simeq \left({N_G^{(m)} \over v_{\cal C}} - {N_G \over N_G^{(m)}} \right)^{-1} \,.
\eeq
Here, the approximation on the right-hand-side neglects the $1/N_G^{(m)}$ term and yields exactly the same result as obtained from the simplified iterative scheme in eq.~(\ref{approximate-recursions}) that ignores the difference between $v_G^{(n)}$ and $\tilde v_G^{(n)}$.

Interestingly, a necessary condition for convergence can be derived also without any approximation directly from eqs.~(\ref{v-G-n-tilde}) and~(\ref{v-G-n}). To this end note that eq.~(\ref{v-G-n}) implies trivially that $v_G^{(n)} \ge \tilde v_G^{(n)}$. Then it follows that
\beq
v_G^{(n)} \le {N_G v_{\cal C} + N_G^{(m)} \over (N_G^{(m)})^2} v_G^{(n-1)} + 
{v_{\cal C} \over N_G^{(m)}} \,.
\eeq
Thus if
\beq
v_{\cal C} \le v_{\cal C}^{crit} \equiv  {N_G^{(m)}(N_G^{(m)}-1) \over N_G}
\eeq
the noise-to-signal ratio $v_G^{(n)}$ converges to a finite value.

%%%%%%%%%%%%%%%%%%
%%% REFERENCES %%%
%%%%%%%%%%%%%%%%%%

\newpage
{\small %footnotesize

} % \end{\small}

%%%%%%%%%%%
% Figures %
%%%%%%%%%%%

\end{document}